\documentclass[twocolumn]{aastex63}
\usepackage{xspace}

\usepackage{threeparttable}

\newcommand{\ncode}[1]{{\sc #1}}
\newcommand{\simba}{\ncode{simba}\xspace}
\newcommand{\arepo}{\ncode{arepo}\xspace}
\newcommand{\caesar}{\ncode{caesar}\xspace}
\newcommand{\cloudy}{\ncode{Cloudy}\xspace}
\newcommand{\python}{\ncode{python}\xspace}

\newcommand{\gizmo}{\ncode{gizmo}\xspace}
\newcommand{\ramses}{\ncode{ramses}\xspace}
\newcommand{\enzo}{\ncode{enzo}\xspace}
\newcommand{\athena}{\ncode{athena}\xspace}
\newcommand{\mappings}{\ncode{MAPPINGS III}\xspace}

\newcommand{\swiftsimio}{\ncode{SWIFTsimIO}\xspace}
\newcommand{\sigame}{\ncode{s\'{i}game}\xspace}
\newcommand{\skirt}{\ncode{Skirt}\xspace}
\newcommand{\yt}{\ncode{yt}\xspace}
\newcommand{\starburst}{\ncode{starburst99}\xspace}

\newcommand{\molH}{\mbox{H$_{\rm 2}$}\xspace}
\newcommand{\hii}{\ion{H}{2}\xspace}
\newcommand{\cii}{[\ion{C}{2}]\xspace}

\newcommand{\nii}{[\ion{N}{2}]\xspace}
\newcommand{\niiA}{[\ion{N}{2}]122\xspace}
\newcommand{\niiB}{[\ion{N}{2}]205\xspace}
\newcommand{\oi}{[\ion{O}{1}]63\xspace}
\newcommand{\oiii}{[\ion{O}{3}]88\xspace}

\newcommand{\LcoA}{\mbox{$L_{\rm CO(1-0)}$}\xspace}
\newcommand{\LcoB}{\mbox{$L_{\rm CO(2-1)}$}\xspace}
\newcommand{\LcoC}{\mbox{$L_{\rm CO(3-2)}$}\xspace}
\newcommand{\Lcii}{\mbox{$L_{\rm [CII]158}$}\xspace}
\newcommand{\Lfir}{\mbox{$L_{\rm FIR}$}\xspace}

\newcommand{\Loiii}{\mbox{$L_{\rm [OIII]88}$}\xspace}
\newcommand{\Loia}{\mbox{$L_{\rm [OI]63}$}\xspace}
\newcommand{\Loib}{\mbox{$L_{\rm [OI]145}$}\xspace}
\newcommand{\Lcia}{\mbox{$L_{\rm [CI]610}$}\xspace}
\newcommand{\Lcib}{\mbox{$L_{\rm [CI]371}$}\xspace}
\newcommand{\LniiA}{\mbox{$L_{\rm [NII]205}$}\xspace}
\newcommand{\LniiB}{\mbox{$L_{\rm [NII]122}$}\xspace}

\newcommand{\Mstar}{\mbox{$M_{\star}$}\xspace}
\newcommand{\Mgas}{\mbox{$M_{\mathrm{gas}}$}\xspace}
\newcommand{\Go}{$G_0$\xspace}
\newcommand{\nH}{\mbox{$n_{\mathrm{H}}$}\xspace}
\newcommand{\NH}{\mbox{$N_{\mathrm{H}}$}\xspace}
\newcommand{\nvw}{$\langle n_{\rm H}\rangle_V$\xspace}
\newcommand{\nSFR}{$\langle n_{\rm SFR}\rangle_V$\xspace}
\newcommand{\Zsfr}{\mbox{$\langle Z\rangle_{\rm SFR}$}\xspace}
\newcommand{\Zmw}{$\langle Z\rangle_{\rm mw}$\xspace}
\newcommand{\FUV}{$F_{\rm FUV}$\xspace}

\newcommand{\cc}{\mbox{cm$^{-3}$}\xspace}
\newcommand{\cs}{\mbox{cm$^{-2}$}\xspace}
\newcommand{\uSFR}{\mbox{M$_{\odot}$~yr$^{-1}$}\xspace}
\newcommand{\unSFR}{\mbox{M$_{\odot}$~yr$^{-1}$kpc$^{-3}$}\xspace}
\newcommand{\Msun}{\mbox{M$_{\odot}$}\xspace}
\newcommand{\Zsun}{\mbox{Z$_{\odot}$}\xspace}
\newcommand{\Lsun}{\mbox{L$_{\odot}$}\xspace}

\newcommand{\nsam}{$400$\,} 
\newcommand{\nsamm}{$240$\,} 

\newcommand{\zsam}{$z=0$\xspace}
\newcommand{\herschel}{{\em Herschel}\xspace}

\usepackage{verbatim}
\usepackage{amsmath}
\usepackage{multirow}
\usepackage{relsize}

\usepackage{amssymb}
\usepackage{float}
\usepackage{hyperref}

\graphicspath{{figs/}}

\newcommand{\sigs}{\sigma_s}

\submitjournal{ApJ}

\shorttitle{SIGAME v3}
\shortauthors{Olsen et al.}

\begin{document}

\title{\sigame V3: GAS FRAGMENTATION IN POST-PROCESSING OF COSMOLOGICAL SIMULATIONS FOR MORE ACCURATE INFRARED LINE EMISSION MODELING}

\correspondingauthor{Karen Pardos Olsen}
\email{karenolsen@arizona.edu}

\author[0000-0003-1250-5287]{Karen Pardos Olsen}
\affiliation{Department of Astronomy and Steward Observatory, University of Arizona, Tucson, AZ 85721, USA}

\author[0000-0001-5817-5944]{Blakesley Burkhart}
 \affiliation{Department of Physics and Astronomy, Rutgers, The State University of New Jersey, 136 Frelinghuysen Rd, Piscataway, NJ 08854, USA}
\affiliation{Center for Computational Astrophysics, Flatiron Institute, 162 Fifth Avenue, New York, NY 10010, USA} 

\author[0000-0003-0064-4060]{Mordecai-Mark Mac Low}
\affiliation{Department of Astrophysics, American Museum of Natural History, New York, NY 10024, USA}
\affiliation{Center for Computational Astrophysics, Flatiron Institute, 162 Fifth Avenue, New York, NY 10010, USA} 

\author[0000-0002-9483-7164]{Robin G. Tre{\ss}}
\affiliation{Universit\"{a}t Heidelberg, Zentrum f\"{u}r Astronomie, Institut f\"{u}r theoretische Astrophysik, Albert-Ueberle-Str. 2, D-69120 Heidelberg, Germany}

\author[0000-0002-2554-1837]{Thomas R.~Greve}
\affiliation{Cosmic Dawn Center (DAWN), Denmark}
\affiliation{DTU-Space, Technical University of Denmark, Elektrovej 327, DK-2800 Kgs. Lyngby, Denmark}
\affiliation{Dept. of Physics \& Astronomy, University College London, Gower Place, London WC1E 6BT, UK}

\author[0000-0001-7610-5544]{David Vizgan}
\affiliation{Astronomy Department and Van Vleck Observatory, Wesleyan University, 96 Foss Hill Drive, Middletown, CT 06459, USA}
\affiliation{Cosmic Dawn Center (DAWN), Denmark}
\affiliation{DTU-Space, Technical University of Denmark, Elektrovej 327, DK-2800 Kgs. Lyngby, Denmark}

\author[0000-0001-7379-2625]{Jay Motka} 
\affiliation{Department of Astronomy and Steward Observatory, University of Arizona, Tucson, AZ 85721, USA}

\author[0000-0002-1327-1921]{Josh Borrow}
\affiliation{Institute for Computational Cosmology, Department of Physics, University of Durham, South Road, Durham DH1 3LE, UK}

\author[0000-0003-1151-4659]{Gerg\"o Popping}
\affiliation{European Southern Observatory, D-85478 Garching bei M\"unchen, Germany}

\author[0000-0003-2842-9434]{Romeel Dav\'e}
\affiliation{Institute for Astronomy, Royal Observatory, Univ. of Edinburgh, Edinburgh EH9 3HJ, UK}
\affiliation{Department of Physics and Astronomy, University of the Western Cape, Robert Sobukwe Road, Bellville 7535, South Africa}
\affiliation{South African Astronomical Observatories, Observatory, Cape Town 7925, South Africa}

\author[0000-0002-0820-1814]{Rowan J. Smith}
\affiliation{Jodrell Bank Centre for Astrophysics, Department of Physics and Astronomy, University of Manchester, Oxford Road, Manchester M13 9PL, UK}

\author[0000-0002-7064-4309]{Desika Narayanan}
\affil{Department of Astronomy, University of Florida, 211 Bryant Space Sciences Center, Gainesville, FL 32611, USA}
\affil{University of Florida Informatics Institute, 432 Newell Drive, CISE Bldg E251, Gainesville, FL 32611, USA}
\affil{Cosmic Dawn Center (DAWN)}

\begin{abstract}
We present an update to the framework called Simulator of Galaxy Millimeter/submillimeter Emission (\sigame). 
\sigame derives line emission in the far-infrared (FIR) for galaxies in particle-based cosmological hydrodynamics simulations by applying radiative transfer and physics recipes via a postprocessing step after completion of the simulation.
In this version, a new technique is developed to model higher gas densities by parameterizing the probability distribution function (PDF) of the gas density in higher-resolution simulations run with the pseudo-Lagrangian, Voronoi mesh code \arepo. The parameterized PDFs are used as a look-up table, and reach higher densities than in previous work.
\sigame v3 is tested on redshift $z = 0$ galaxies drawn from the \simba cosmological simulation for eight FIR emission lines tracing vastly different phases of the interstellar medium. 
This version of \sigame includes dust radiative transfer with \skirt and high resolution photo-ionization models with \cloudy, the latter sampled according to the density PDF of the \arepo simulations to augment the densities in the cosmological simulation. 
The quartile distributions of the predicted line luminosities overlap with the observed range for nearby galaxies of similar star formation rate (SFR) for all but two emission lines: \oi and CO(3-2) which are overestimated by median factors of $1.3$ and $1.0$\,dex, respectively, compared to the observed line--SFR relation of mixed-type galaxies.
We attribute the remaining disagreement with observations to the lack of precise attenuation of the interstellar light on sub-grid scales ($\lesssim200\,$pc) and differences in sample selection.
\end{abstract}


\section{Introduction} \label{sec:intro}


The physical and chemical state of the interstellar medium (ISM) plays a key role in determining the star formation rate (SFR) of a galaxy and is therefore of immense importance for understanding galaxy evolution. 
Macroscopic galactic events such as starbursts and mergers create pressure and density waves that determine where and how the giant molecular clouds (GMCs) form in which stars and planets grow
\citep[e.g.][]{colombo2014,sun2018,chevance2020,alves2020}. 


Resolved observations of nearby clouds and detailed simulations that track individual cores in GMCs from initial perturbations to final collapse can probe the actual density distribution above the highest density achievable in cosmological simulations. 
Both observations and numerical work to this effect have determined that the dense gas in molecular clouds (with extinctions $A_v >1$ and/or $n > 10^3$~cm$^{-3}$) typically has a density probability distribution function (PDF) with a power law tail at the high end set by gravitational collapse \citep{Klessen2000,Kainulainen2009,Collins12a,Girichidis2014,Burkhart2015,schneider2015MNRAS.453L..41S,Lombardi2015AA,burkhart2017,Mocz2017,padoan2017ApJ...840...48P,Alves2017AA,Chen2018}. 
The low-density portion of the density PDF might follow a lognormal shape but this is highly uncertain due to selection effects, completeness, and the influence of stellar feedback \citep{Alves2017AA,Chen2018}. 

On galactic scales, it becomes too computationally expensive for simulations to model the collapse and fragmentation of each cloud, not to mention the subsequent stellar feedback processes. Instead, parameterizations are used, for example in the cosmological simulations  \simba \citep{dave2019} and IllustrisTNG \citep{weinberger2017,pillepich2018}. In \simba, an \molH-based SFR recipe is used where the amount of \molH is estimated from the metallicity and local column density following the sub-grid model of \cite{krumholz2011}. In IllustrisTNG, gas with density $n \gtrsim0.1\,$\cc is allowed to form stars in accordance with the empirically defined Kennicutt-Schmidt relation. Both simulations lack the resolution to track individual stars and instead assume a \citet{Chabrier2003} initial mass function (IMF) for stellar populations formed.


Synthetic observations of simulated galaxies are our most promising way of directly comparing models with observations. 
Observations of the ISM in galaxies typically target its most effective cooling channels, namely far-infrared (FIR) emission lines.  These were discovered by 
the Kuiper Airborne Observatory \citep{Watson1984,Stacey1991,Lord1996}. 
The subsequent Infrared Space Telescope \citep{Kessler1996} and Herschel Space Observatory \citep[\herschel hereafter; ][]{Pilbratt2010} showed that the strongest emission lines come from a handful of species, namely neutral and ionized oxygen, singly ionized carbon, and ionized nitrogen, as well as molecular lines such as the CO rotational lines. Together, these species probe all phases of the ISM from GMCs to photon-dominated regions (or photodissociation regions; PDRs), warm neutral medium and hot ionized gas. Existing telescopes such as ALMA, NOEMA, and SOFIA as well as upcoming space and balloon missions (e.g.\ GUSTO and ASTHROS) provide an ever-growing database of emission line observations of galaxies at all redshifts, requiring detailed modeling of the same lines for their correct interpretation. 
By taking a snapshot of a hydrodynamic simulation containing gas, stars, and dark matter, line emission can be calculated as a postprocessing step for each cell or particle in the simulation using physically motivated recipes. 

All efforts to simulate emission lines have had to determine small-scale values for:
1) the structure and local spectral shape of the radiation field, in particular in the far-ultraviolet (FUV), 2) the structure of gas density, and 3) the local chemistry and level populations. Due to computational constraints, most of these are not calculated with enough precision in the simulation itself, and hence must be derived in postprocessing based on sub-grid models.
Below, we summarize the current approaches:
\begin{enumerate}

    \item For the FUV, the current practice typically involves adopting the strength \Go of FUV interstellar radiation field (ISRF) in the solar neighborhood \citep{habing1968} and scaling it by the global SFR of the simulated galaxy compared to that of the Milky Way \citep[e.g.][]{Vallini2019} to get a galaxy-averaged FUV flux, or by using the local SFR density to get a local kiloparsec-scale FUV flux \citep[e.g.][]{Olsen17a,Popping2019}.  \citet{Pallottini2019} demonstrated the use of on-the-fly radiative transfer for a high-resolution adaptive mesh refinement simulation of a $z\sim6$ galaxy, and some large cosmological simulations have now also seen on-the-fly radiative transfer, albeit only from Big Bang through the epoch of reionization and not down to low redshifts \citep{Ocvirk2016,Wu2019,Ocvirk2020}. {\it In this work, we introduce a postprocessing radiative transfer calculation for large-box simulations at $z=0$}
    
    \item Retrieving the high-density ($n>10^4$\,\cc) gas during postprocessing of a cosmological simulation is crucial to modeling several FIR lines, as can be seen from the critical densities of typical FIR lines listed in Table\,\ref{tab:crit}. This step is typically done by adopting a clumping factor that artificially increases the H$_2$ production \citep[e.g.][]{Narayanan2008, Gnedin2009,Dave2016,Lupi2020}, adopting a locally observed molecular cloud mass spectrum together with a derived cloud radius and assumed inner density profile \citep[e.g. ][]{Olsen15a,Olsen16,Olsen17a,Popping2019,Inoue2020}, or fragmenting the gas on sub-grid scales following a lognormal density PDF as motivated by \citet{Padoan2011} for isothermal molecular clouds, either using a fixed Mach number \citep{Vallini2019,leroy2017} or calculating the Mach number from local gas properties \citep[e.g.][]{Narayanan2014,Pallottini2019}. However, the use of the lognormal PDF may be physically unmotivated as evidenced by the lognormal+power law PDF seen in observations and simulations, as described earlier. \cite{Vallini2018} adopted a lognormal+power law PDF for their modeling of CO lines, and we will adopt a similar formalism. {\it In this work, we employ simulations of higher spatial resolution as look-up tables and use analytic relations of density PDFs from \citet{Burkhart2018} to infer the PDF on sub-grid scales in a cosmological simulation.}
    
    \item Once the radiation field and density are determined, there are many tools publicly available that simultaneously solve for ionization state, temperature, and line excitation, with some of them being tailored to a specific ISM phase. For an overview of widely used tools to solve for the chemistry and generate line emission in the ISM, see \citet{Olsen2018}, which also discusses the limitations of each approach. Most importantly, the chemistry and emission can be solved
    the simulation or, for a more precise result, with nonequilibrium on-the-fly techniques but at greater computational cost. For an in-depth comparison of the two methods in the case of \cii and \oi see \cite{Lupi2020b}. {\it In this work, we use the photoionization code \cloudy \citep{Ferland17a} to postprocess the simulations to set the shape of the spectrum locally as well as to calculate chemistry and line emission.}
\end{enumerate}

As previously mentioned, in addition to inferring the PDF in higher-resolution simulations as described in point 2 above, we also use a theoretical framework to estimate the amount of self-gravitating gas not resolved in the galaxy-sized simulation. 
In order to alleviate tensions with observations and dense gas models that use a lognormal PDF,  \citet{2019ApJ...879..129B} derive a model of dense gas and star formation based primarily on the presence of a power law tail. 
The work provides an analytic expression connecting the transitional column density value between lognormal and power law to the width and of the lognormal and the slope of the power law which we will use here \citep[see also][]{Collins12a,burkhart2017,Burkhart2018}.  

\begin{deluxetable}{lp{2.8cm}p{3cm}}
\tablecaption{Critical Densities of Important ISM Cooling Lines in the FIR.\label{tab:crit}}
\tablehead{\colhead{Line}&\colhead{$n_{\mathrm{crit}}$ (cm$^{-3}$)}&\colhead{Origin}}
\startdata
\cii            &   16\tablenotemark{a},  $2.4\times10^3$\tablenotemark{b},$4.8\times10^3$\tablenotemark{c}   & All ISM, but mainly PDRs\\
\oi             &   $4.7\times10^5$\tablenotemark{b} & PDRs    \\
\oiii           &   $510$\tablenotemark{a}   & \hii regions, radiation dominated by OB stars\\
\niiA           &   $310$\tablenotemark{a}      & Ionized ISM \\
\niiB           &   $48$\tablenotemark{a}       & Ionized ISM\\
CO(1-0)         &   $650$\tablenotemark{c}      & Molecular ISM\\
CO(2-1)         &   $6.2\times10^3$\tablenotemark{c}      & Molecular ISM\\
CO(3-2)         &   $2.2\times10^4$\tablenotemark{c}      & Molecular ISM
\enddata
\tablecomments{CO critical densities were calculated using the spontaneous emission coefficients from the LAMDA database as accessed on 21 Aug 2020 \citep{Schoier2005}, using a temperature of $100$\,K and typical collision cross section of $10^{-15}$~\cc. The \cii critical densities are from \citet{goldsmith2012}, calculated at a temperature of $500\,$K. The remaining critical densities are from \citet{Madden2013} at $300$\,K.}
\tablenotetext{a}{For collisions with electrons.}
\tablenotetext{b}{For collisions with hydrogen atoms.}
\tablenotetext{c}{For collisions with H2 molecules.}
\end{deluxetable}


With this paper, we present a new version of our postprocessing framework Simulator of Galaxy Millimeter/submillimeter Emission \citep[\sigame;][]{Olsen17a} to model FIR/millimeter line emission in particle-based hydrodynamic simulations, and test it on a large sample of emission lines for comparison with observations. Section \ref{sec:sigame} describes the new structure of \sigame in detail and lists the updates made since the previous version. 
Section\,\ref{sec:sample} describes the test sample of simulated galaxies used, and the rest of Section\,\ref{sec:val} tests the code under different assumptions and makes a rough comparison with observations and the previous version of \sigame. 
Section\,\ref{sec:dis} discusses the results and provides caveats for using this method. 
Finally, we conclude in Section\,\ref{sec:con}.

Throughout this paper, we adopt a flat cold dark matter cosmology with cosmological parameters $\Omega_{\Lambda}=0.693$, $\Omega_m=0.307$, $\Omega_b=0.048$, and dimensionless Hubble parameter $h=0.678$ \citep{Planck16a}.

\begin{figure*}[htbp]
\centering
\includegraphics[width=.6\textwidth]{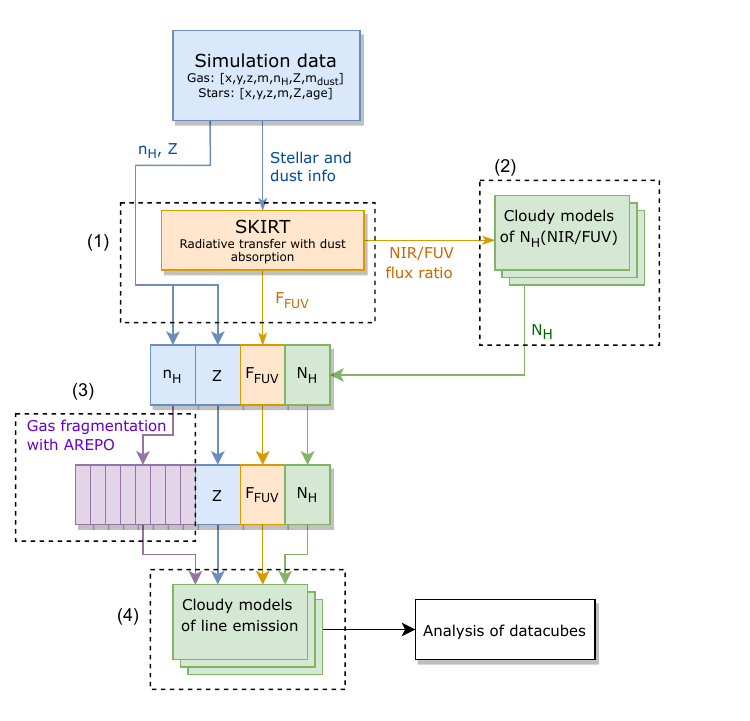}
\caption{Flow chart illustrating the structure of SIGAME v3. \label{fig:chart}}
\end{figure*}

\section{S{\'I}GAME version 3} \label{sec:sigame}
\sigame derives FIR/millimeter line emission of simulated galaxies by postprocessing a particle-based cosmological hydrodynamics simulation. 
We currently use particle-based simulations, but grid-based simulations are in principle also usable with the new framework presented here. 
Figure\,\ref{fig:chart} provides a quick look at the overall structure of \sigame version 3. 
\cite{Leung2020} adapted version 2 of \sigame to apply it to the \simba simulation for the first time to study the \cii luminosity function at $z=6$. With this version 3, the main updates are:



\begin{itemize}
    \item Radiative transfer of stellar FUV emission through dust. \sigame v2 used a library of \starburst \citep{Leitherer2014} synthesis models for stellar populations to derive the local FUV field, without taking into account the potentially important effect of dust absorption. In \sigame v3 a full dust radiative transfer calculation is performed using the three-dimensional (3D) radiative transfer code \skirt\footnote{\url{http://www.skirt.ugent.be/}} \citep{Camps2020} and the actual dust-to-metal (DTM) ratios from the cosmological simulation simulation if available. 
    
    \item Fragmentation of the gas on sub-grid scales. In previous versions of \sigame, gas particles from cosmological simulations were divided into a diffuse gas phase and a dense one, the latter being further divided into spherical GMC-like structures with masses $>10^4\,\Msun$. With the new version of \sigame, gas particles are first distributed on an adaptive grid with \skirt and then fragmented according to the local gas and SFR volume densities using the output from a high resolution simulation, and thereby avoiding the assumption of spherical clouds. 
    
    \item High-resolution thermochemistry of the ISM. As in previous versions of \sigame, we use the photoionization code \cloudy (this time version 17.02) to calculate the chemical and thermal state of the gas and the resulting line emission. However, because of the decision to fragment the gas to smaller mass and size scales, we have moved to using the \ncode{grid command} feature of \cloudy, allowing for a look-up table of higher resolution.
\end{itemize}

The following four subsections describe in detail the algorithms that are applied within each of the four steps indicated in Figure\,\ref{fig:chart}:

\subsection{Dust attenuation with \skirt (step 1)} \label{sec:step1}
We use the 3D radiative transfer code \skirt \citep{Camps2020} to calculate the local dust-attenuated ISRF. 
\skirt uses information on the dust and stellar distribution in the galaxy simulation to inform a Monte Carlo algorithm on how to emulate the relevant physical processes including scattering, absorption, and emission of photons as they pass through the interstellar dust. 
For the stellar component (``SourceSystem" in \skirt), an IMF similar to that used in the galaxy simulation is chosen, and the stellar emission is set to match the Binary Population and Spectral Synthesis (BPASS v2.2.1) models according to the age and metallicity of each star particle \citep{Eldridge2017,Stanway2018}.

\skirt requires radiative smoothing lengths around each stellar particle's position to spread out the launch locations of the photon packets around that position. If such a smoothing length is not available in the simulation, \sigame will calculate one that scales linearly with stellar mass and spans 100--300\,pc. 
Often, \skirt divides the dust medium into cells of sizes much smaller than 100\,pc, but we note that the gravitational smoothing lengths of cosmological simulations are typically much larger (of the order of 1\,kpc). Therefore, although SKIRT inserts a lot of cells to ease the radiative transfer calculation, adjacent cells often have similar densities. 
The choice of scaling the radiative smoothing length to between 100 and 300\,pc was made after having tested two extreme cases: (a) stellar smoothing lengths of 10\,pc and (b) stellar smoothing lengths of 1000\,pc. In case (a) the stellar radiation field ends up being concentrated in point sources and in the case (b) the stellar radiation field creates artificial ``rings'' around the original stellar particle position. Hence, we chose the smoothing lengths somewhere in between. \footnote{For more detail, we direct the reader to an issue on the \skirt GitHub site \citep{skirt_issue}, in which we discussed this question at length with \skirt developer Peter Camps, and where plots can be found of the various options tested.}

Finally, \skirt requires the initial mass of each stellar particle rather than the current stellar mass, since the spectral energy distributions used by \skirt already take into account the mass evolution of the population based on its age and metallicity. If not available in the simulation, \sigame will calculate the initial stellar mass, before mass loss due to stellar evolution, using the publicly available Python-FSPS code, which itself is a Python translation of the Flexible Stellar Population Synthesis code, in order to convert current stellar mass and formation time into initial stellar masses \citep{Conroy2009,Conroy2010,fsps}. 
For the dust component (``MediumSystem'' in \skirt), \sigame can either use the dust masses directly, if provided by the galaxy simulation, or calculate them from the metallicity of the gas using a fixed DTM mass ratio.
For the dust types and composition, we use the built-in THEMIS model \citep{Jones2017} for dust in the diffuse ISM that can be invoked in \skirt. This dust mixture contains two families of dust particles: amorphous silicate and amorphous hydrocarbon (we refer to \citealt{Jones2017} for more detail). 
The postprocessing with \skirt returns a spectrum for each cell of each galaxy on an oct-tree adaptive grid. 
The rest of \sigame works on this grid structure, for which gas densities are calculated with the \swiftsimio package \citep{Borrow2020}, and all other properties are inherited from the nearest fluid element or gas particle in the galaxy simulation.

\begin{figure*}[t]
\centering
\includegraphics[width=.8\textwidth]{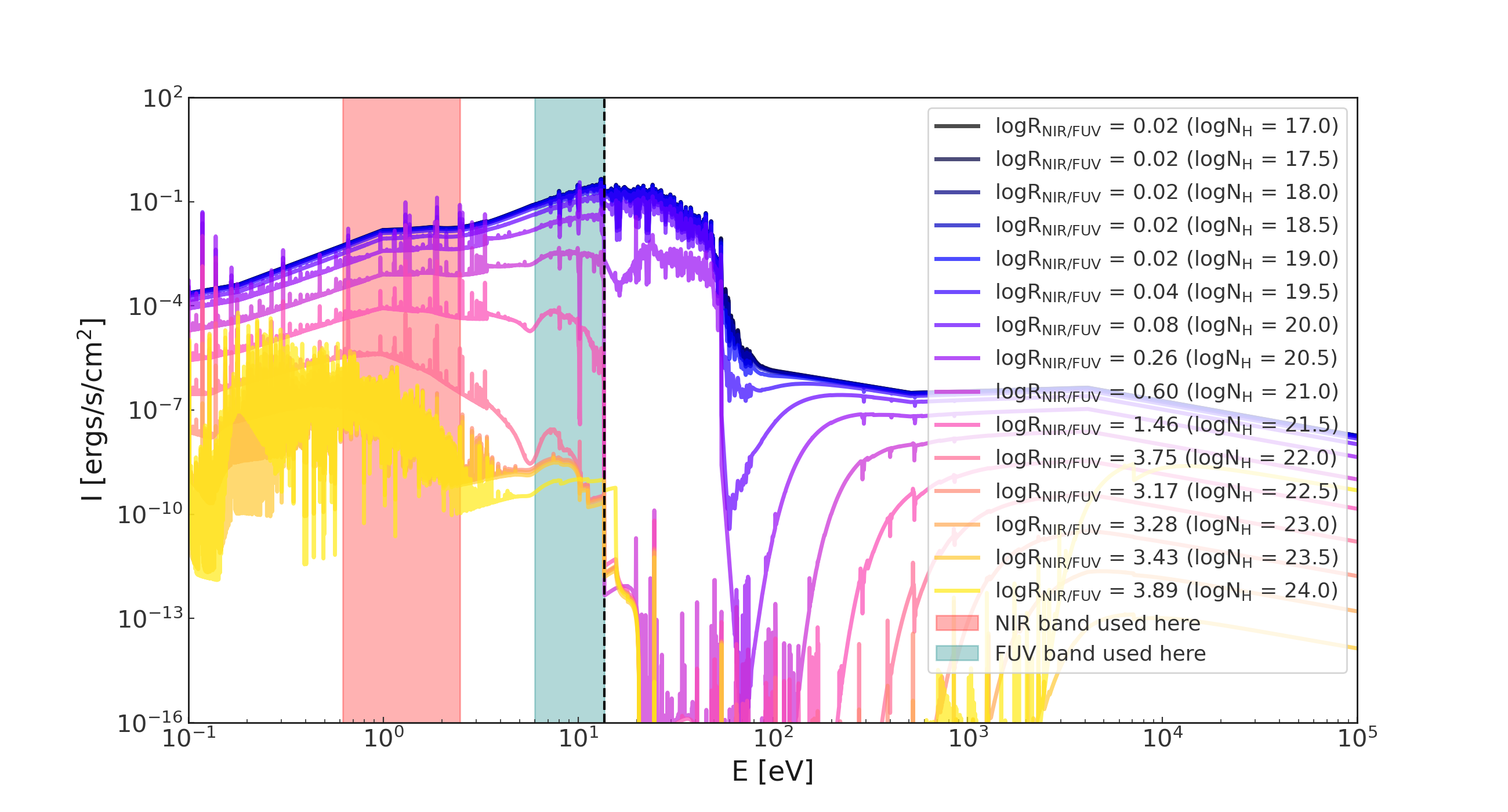}
\caption{Transmitted spectra for a set of \cloudy~ 1D slab models with hydrogen density $1\,$\cc, and solar dust and metal abundances. The OIR and FUV photometric bands that we use to derive the OIR/FUV flux ratio are indicated with shaded regions, and a vertical dashed line highlights the 13.6\,eV ionization potential of hydrogen. 
The luminosity source used is a combination of the local ISRF continuum together with the spectrum of a BPASS v2.2.1 binary star model with age $10^6$\,Myr and solar metallicity and with a bolometric luminosity normalized to $10^6$\,\Lsun. 
The different spectra were created by changing the column density of the 1D slab from $\log\,N_{\rm H} = 17$ to $\log\,N_{\rm H} = 24$.
\label{fig:cloudy_NH}}
\end{figure*}

\subsection{Attenuation by intervening gas with \cloudy (step 2)}\label{sec:step2}
In addition to being attenuated by dust grains, interstellar light from stars is also absorbed by gas, in particular in the hydrogen-ionizing regime at photon energies above $13.6$\,eV. With \skirt a spectrum is generated in each cell that has been attenuated by dust only, which tends to absorb radiation in the FUV at $6$--$13.6$\,eV \citep{Gnedin2008}. 
In order to account for the additional attenuation by gas, we run a suite of \cloudy models for a fixed luminosity source, over a range of total hydrogen column densities. 
For the luminosity source, an unobscured stellar spectrum of a stellar population of age 1\,Myr and solar metallicity, scaled to a luminosity of $10^6$\,\Lsun, using the same BPASS v2.2.1 tables as those used in the modeling with \skirt. 
The stellar spectrum is added to a background continuum spectrum corresponding to the observed local ISM ISRF as available with the ``table ism'' command in \cloudy. 
For these models, we keep the metallicity at $1$\,\Zsun and maintain the dust size distribution and abundance to that appropriate for the ISM of the Milky Way as given by the \cloudy ``grains ISM'' command. 
The latter includes both a graphitic and a silicate component and generally reproduces the observed overall extinction properties for a ratio of extinction per reddening of RV $\equiv$ Av/E(B-V) = 3.1 (we refer to the \cloudy manual for details). 
We use the local ISM abundance table of \cloudy, which is a mean of the warm and cold phases of the ISM from \cite{Cowie1986} and \cite{Savage1996}. 
In these \cloudy models, and all other \cloudy models in this work, the cosmic microwave background at $z=0$ is included as an additional blackbody radiation field. 
By changing the column density at which \cloudy stops the calculation, a set of transmitted spectra of different shapes are produced containing the effects of different amounts of dust and gas absorption as shown in Figure\,\ref{fig:cloudy_NH}. 

From the shape of the spectra in Figure\,\ref{fig:cloudy_NH}, we quantify the effect of dust absorption as the ratio between optical-to-near-infrared (OIR; 0.5--2\,eV) flux and FUV flux (red and teal shaded regions in Fig.\,\ref{fig:cloudy_NH}). 
We can now match the derived OIR/FUV flux ratio with that of \skirt in each simulation cell to translate the dust extinction into an average column density of gas that the stellar light has passed through, and thereby complete the shape of the incident spectrum (not the normalization) to be used for the final \cloudy grid described in Section\,\ref{sec:step4}. 
The inclusion of absorption by gas in this way is not an exact solution because we are essentially approximating the effect of several luminosity sources and column densities taken into account by \skirt with a single luminosity source and one column of gas and dust in \cloudy. We do this because there is no clear way of recovering the exact path of each photon packet ejected by \skirt. Furthermore, this method assumes a constant DTM ratio (fixed at roughly $0.5$ here) and uses a single metallicity for all the intervening gas. Finally, we note that the chosen stellar population sets the X-ray portion of the spectra in \cloudy which is further attenuated by gas and dust as seen in Fig.\,\ref{fig:cloudy_NH}. Yet this part of the spectrum could look very different for a stellar population of different age and metallicity, and the intensity of X-rays could potentially affect the emission line flux \citep[see ][ for a study on CO emission from high-$z$ active galactic nuclei (AGNs)]{Vallini2018}.

\subsection{Gas fragmentation (step 3)}\label{sec:step3}
Due to the limited resolution of cosmological simulations, the individual cells typically do not contain densities higher than $\sim10$\,\cc (see Appendix\,\ref{app:cell} for the gas and SFR density distributions in one of the simulated galaxy samples used to test \sigame in the following sections). 
However, all the FIR emission lines considered here can or will only be excited at much higher densities, as can be seen from the critical densities listed in Table\,\ref{tab:crit}. 
In order to perform the dust radiative transfer, \skirt introduces cells of smaller size than those in the source simulation, but this procedure does not increase the density of the gas. 

To mitigate the lack of resolution in density, this version of \sigame turns toward simulations made at much higher spatial resolution in order to use them as look-up tables that can help describe the fragmentation of the gas on sub-grid scales. In principle, the user can import their simulation of choice, but for the purpose of testing \sigame in this paper we chose the high-resolution data from a simulation performed with the Voronoi mesh
code \arepo \citep{Springel2010} of an interacting M51-like galaxy model \citep{tress2020}. The simulation (hereafter described as \arepo-M51) reached subparsec resolution in dense gas and allowed for an analysis of the formation and destruction of GMCs, which showed that the evolution of GMCs depends only weakly on galaxy--galaxy interactions.  

\subsubsection{The \arepo-M51 simulation}

For details on the simulation setup we refer to \citet{tress2020}, summarizing here only the components of relevance to this project. The \arepo-M51 simulation was designed to be able to resolve and study GMCs in the context of a galaxy interaction. The simulation setup includes a time-dependent, nonequilibrium, chemistry network tracking hydrogen and CO chemistry, and follows star formation and feedback processes, reaching subparsec resolutions at densities $n \gtrsim100$~\cc (see Fig.~3 of \citealt{tress2020}). The calculations were executed using \arepo, a magnetohydrodynamic (MHD) code using an approximate Riemann solver coupled to an oct-tree gravity solver, from which only the hydrodynamic capabilities were used \citep{Springel2010,Weinberger2020}. At each time step the code constructs the Voronoi grid given a set of mesh-generating points that are constrained to move following the local velocity of the fluid. This pseudo-Lagrangian
moving-mesh technique is naturally adaptive, allowing for high dynamic range on spatial scales down to the substructure of GMCs.

The galaxy model and the interaction with a companion were adjusted to resemble the M51 system, and the chemical network followed the cooling and self-shielding of molecular gas from foreground interstellar radiation. This allowed the formation of GMCs where runaway collapse leads to star formation (SF). Dense, gravitationally bound, and collapsing gas is accreted onto collisionless sink particles that abstract the last stages of the SF process to a sub-grid model due to the limited resolution of even this higher-resolution simulation.

At gas densities $\rho_{\rm c} > 10^{-21}$\,g\,\cc ($n_{\rm H}\gtrsim 600$\,\cc) the gas is tested to determine whether it is bound and collapsing, and if so a sink particle is created. Dense gas with $\rho > \rho_{\rm c}$ that comes within an accretion radius of 2.5\,pc will then be accreted by the sink particle if bound to it. Given these densities and sizes, the sink particles represent small stellar (sub)clusters. At these scales SF is still fairly inefficient, so only $5$\,\% of the accreted gas mass is converted into stars. In this sense every sink particle consists of a stellar component and a gas component. Each stellar component is modeled as a stellar population whose initial stellar masses are drawn from an input \citet{Kroupa2001} IMF. The massive stars drawn are evolved based on a simple stellar evolution model \citet{Maeder2009} and at the end of their lifetime they produce a supernova (SN) feedback event, disrupting the clouds and closing the ISM lifecycle. Note that this approach is different from the traditional star particle approach of cosmological simulations, as we explicitly consider the gravitational binding of the gas to model ISM fragmentation and then include subsequent gas accretion.

The gas component trapped within the sink particle is progressively ejected and re-added to the gas phase with every SN event. After the last SN of a particular sink, the sink particle includes only stellar mass and is converted to a different particle type that represents an old stellar population. By drawing randomly using Poisson sampling from the IMF, it could occur that in particular cases no massive star is generated within the sink. In that case, after a trial period of $10$\,Myr, the gas component is ejected from the sink without an SN event and the sink is then converted to an old stellar population particle. Significant approximations of the \arepo-M51 simulation include the lack of early stellar feedback such as winds, jets, and ionizing radiation and the absence of magnetic fields.

\begin{figure*}[htbp]
\centering
\includegraphics[width=1\textwidth]{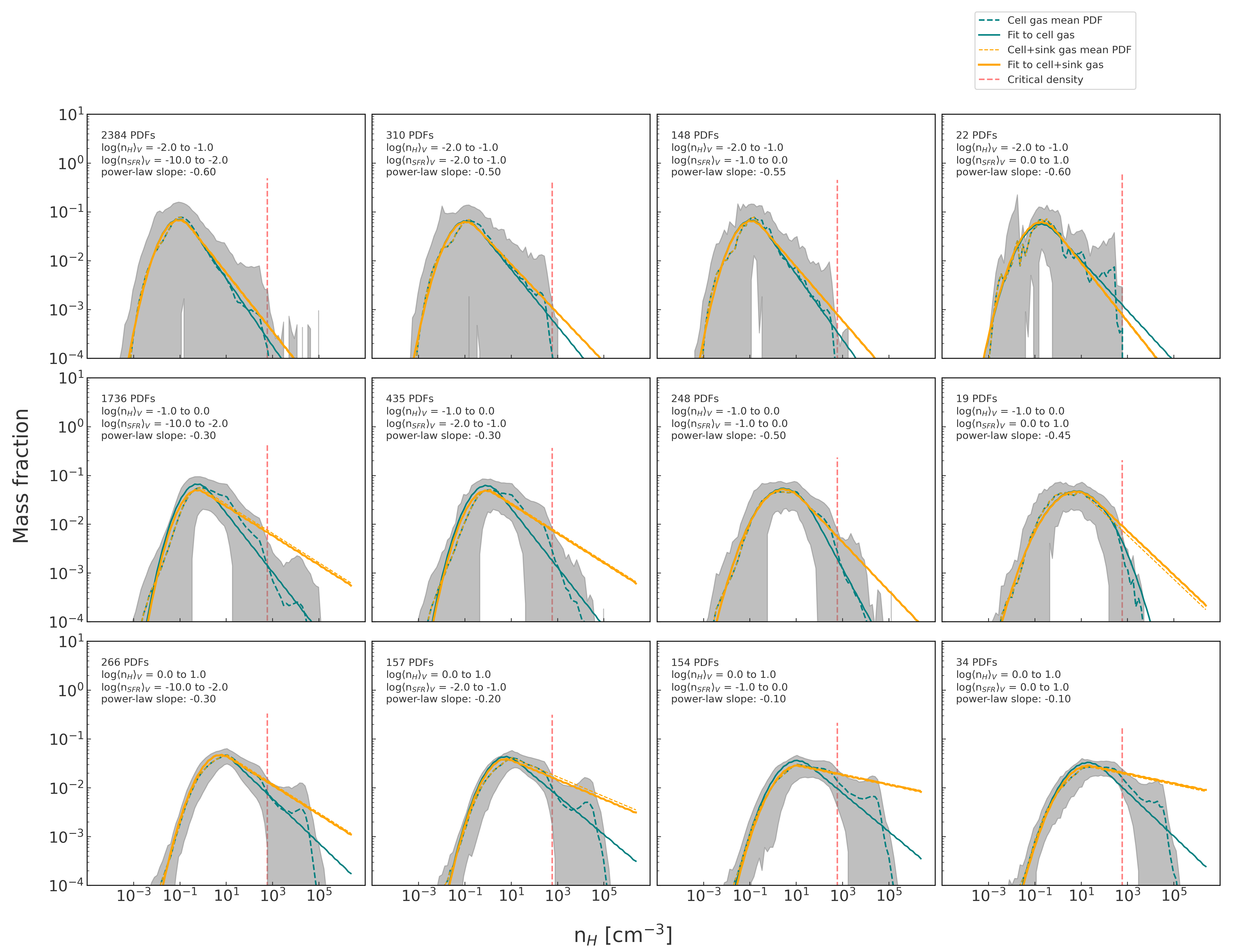}
\caption{Volume-weighted gas density PDFs from the selected \arepo-M51 run and fitted functions constructed as described in Section\,\ref{sec:step3}. Each panel indicates a specific bin in gas density, \nvw, and SFR volume density, \nSFR. The gray contours correspond to the 1-$\sigma$ spread around the mean volume-weighted PDF (dashed teal curve) made from AREPO gas cells within regions of $(200\,$pc$)^3$. 
 The solid teal curves are lognormal+power law fits to the mean PDF. A vertical dashed red line indicates the critical density above which sink particles may form, at which point the density PDF from gas cells alone is no longer comprehensive. The orange dashed curves account for the dense gas in sink particles by adding this mass at the high densities in the form of a modified power law slope, and the orange solid curves are the resulting lognormal+power law fit to the new distribution. 
 The units for \nvw and \nSFR are in \cc and \unSFR, respectively.
 \label{fig:AREPO_PDFs}}
\end{figure*}

\subsubsection{Parameterizing the density PDF}

The \arepo-M51 simulation contains starbursting regions as well as regions with hardly any ongoing star formation. The question we pose is: how does the gas density PDF change from one region to another and can we parameterize it for use in a cosmological simulation of much lower resolution? 
To investigate this, we divide the \arepo-M51 simulation volume into cubes of $200$\,pc on a side and calculate the gas density PDF within each cube. 
The region size of $200$\,pc was chosen in order to represent typical \skirt cell sizes, and at the same time be large enough to contain enough elements in \arepo-M51 to properly sample the gas density PDF. We expect the PDF to move toward higher densities as the volume-averaged gas density of a region increases, but also as SFR density increases \citep{Kainulainen2009}. Figure\,\ref{fig:AREPO_PDFs} shows the resulting mean PDFs of cells within $200$\,pc regions using teal dashed lines. In gray we show the $1\sigma$ spread for 12 bins of volume-averaged density, \nvw, and SFR density, \nSFR. A red vertical dashed line in each panel indicates the critical density, above which the cell data start to be incomplete as some gas cells are converted into sink particles.  
 
Due to the spread in PDFs, as indicated with gray areas in Figure\,\ref{fig:AREPO_PDFs}, we cannot just interpolate between PDFs. Instead, we make a parametric fit, such that the fit parameters depend on \nvw and \nSFR. For the parametric fit, we build on previous work showing that the density distribution of a collapsing cloud can be approximated by a lognormal with a power law tail.
As discussed in the Introduction, both observations and simulations show
that the density field of an isothermal star-forming cloud is well approximated by a lognormal distribution at low density and a power law distribution at high density. \citet{Burkhart2018} describes the resulting PDF as
\begin{align}
p_{LN+PL}(s) = 
\begin{cases}
 N\frac{1}{\sqrt{2\pi}\sigs}e^{\frac{( s - s_0)^2}{2\sigs^2}}, & s < s_t \\
 N  e^{-\alpha s}, & s > s_t ,
\end{cases}
\label{eqn.piecewise}
\end{align}
where $N$ is the normalization constant, $\sigs$ is the standard deviation of the lognormal, $\alpha$ is the power law slope, $s_0$ is the mean logarithmic overdensity, and $s$ is the logarithmic overdensity 
\begin{equation}
    s\equiv \ln(\rho/\rho_0),
\end{equation}
where $\rho_0$ is the volume-averaged density of the entire cloud. The transition density, $s_t$, above which the distribution approximates a power law, was shown by \cite{Burkhart2018} to relate to the lognormal width $\sigma_s$ and power law slope $\alpha$ as
\begin{equation}
    s_t = (\alpha-1/2)\sigma_s^2.
\end{equation}
In Figure\,\ref{fig:AREPO_PDFs} a fit is made to all mean PDFs by combining a lognormal function with a power law at the high-density end, keeping $\sigma_s$ and $\alpha$ as free parameters.

However, this fit does not take into account gas mass locked in sink particles. The gas in sink particles where no SN has exploded will be dense and distributed in relatively undisturbed GMCs, while gas in other sink particles with supernova remnants (SNRs) has already been at least partly dispersed in the surrounding ISM due to SN feedback. In order to account for the dense gas mass in sink particles, we divide the sink particles into two categories:

\begin{itemize}
\item Category I sink particles contain dense, cold gas that creates stellar populations with time, but no SN explosions. Here, we expect the gas to be relatively undisturbed and we count all gas mass in the particle as dense gas following a power law density PDF regardless of the sink particle age. The percentage of sink particles in this category is only $0.13\,\%$.
\item Category II sink particles contain dense, cold gas that has formed or will produce at least one SN. Due to feedback from the massive stars that eventually produce SN explosions, a GMC will not survive for long in this environment. In a study looking for OH($1720$\,MHz) masers in the inner Galaxy, \cite{Hewitt2009} found that dense gas 
interacts with only 15\% of SNRs. Considering that SNRs are visible only during the first $\sim0.1\,$Myr of their lifetime and assuming that SNe occur at a roughly constant rate after the first SNe go off, we can convert the SNR number fraction of $15\%$ into an upper age limit. We count gas in sink particles with SNRs as being dense gas only if the sink particle age since the first SN exploded is less than 15\% of the longest SNR lifetime. The first SNe, created by the most massive stars, are expected to begin exploding at a sink particle age of $\sim3\,$Myr \citep[e.g. ][]{Yungelson2008}. The percentage of sink particles in this category is $22.7\,\%$.
\end{itemize}

\begin{figure}[htbp]
\centering
\includegraphics[width=.5\textwidth]{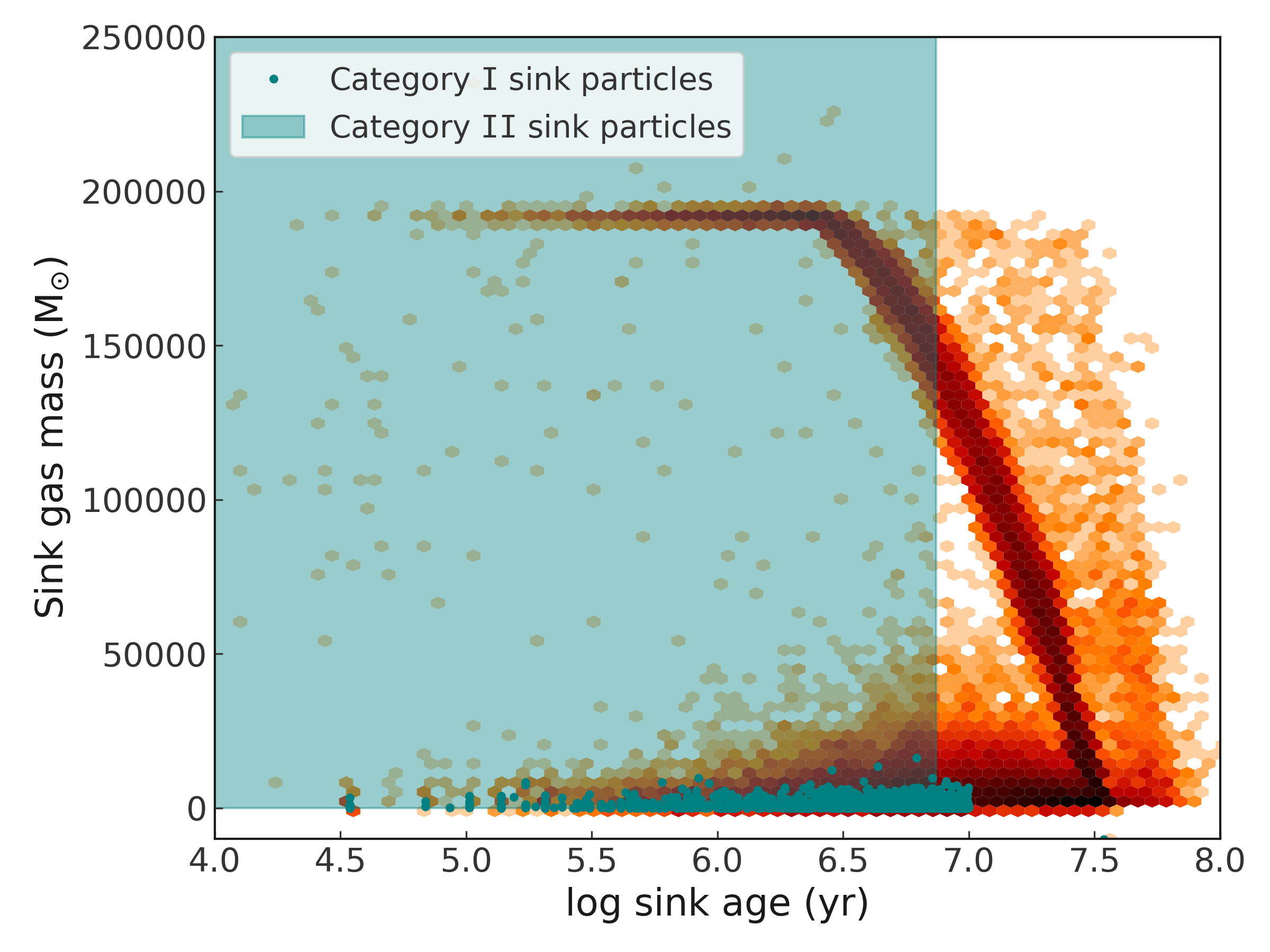}
\caption{Distribution of sink particle gas mass and age in \arepo-M51. The teal circles and teal shaded area illustrate the selection criteria of Category I and II sink particles, respectively, as described in Section\,\ref{sec:step3}. 
\label{fig:sinks}}
\end{figure}

We count the internal gas mass of the remaining sink particles as diffuse gas that would already have been dispersed by feedback if ionization and winds were included. As sink particles accrete more mass than forms into stars, there is a maximum mass limit of  $\sim2 \times 10^5$\,\Msun above which they are not allowed to accrete further and instead form a new sink particle. This retains nominal resolution in the collisionless particles instead of having single particles with anomalously high mass. 
Figure\,\ref{fig:sinks} illustrates the selection criteria and the distribution in sink particle ages and total current gas masses (total sink mass -- sink stellar mass). The older the sink particles, the more likely it is that stars have formed and the gas has been used up or dispersed. This outweighs the additional mass from accretion over time. 
A main branch can be seen of sink particles that start out with the maximum mass and begin to reduce in gas mass after a period of 3~Myr when SNe start to disperse the gas (Category II sink particles by the definition above).
Another branch is of sink particles that ran out of gas, many of which are Category I particles that will never produce an SN.

The additional sink gas mass is expected to change only the high-density portion of each PDF, because it has already passed the accretion density threshold of $\rho_c = 600$\,\cc.  We first attempt to accommodate the sink gas mass by iterating on the power law slope until enough additional mass has been added. 
This method results in the orange dashed curves in Figure\,\ref{fig:AREPO_PDFs}, while the orange solid curves are combined lognormal and power law fits made to the new distributions. 

The resulting fits, shown with orange curves in Figure\,\ref{fig:AREPO_PDFs}, are used as look-up tables for each cell in the \skirt output, by interpolating in \nvw and \nSFR. 
The corresponding sonic Mach numbers, $M_s$, for each fit can be estimated from $\sigs$ through \citep{Padoan1997};
\begin{equation}
    \sigs^2 = \ln(1 + b^2M_s^2).\label{eq:Mach}
\end{equation}
Adopting a turbulent forcing parameter of $b = 1/3$, our density PDFs display Mach numbers from 7 to 27. 
For cell densities below the minimum grid point in the \arepo-M51 look-up tables, we adopt a single, narrow lognormal PDF corresponding to a Mach number of $1$ and $b = 1/3$ \citep{Federrath2008} without the power law tail. All PDFs are cropped to the density range from $\nH=10^{-4}$\,\cc to $\nH=10^{7}$\,\cc.

\subsection{FIR line emission with \cloudy (step 4)}\label{sec:step4}
The final step of \sigame v3 is to derive the line emission, having gathered all the necessary information by postprocessing the simulation. To this end, we use a library of one-zone \cloudy models that span the parameter ranges listed in Table\,\ref{tab:cloudy_param}. As mentioned in Section\,\ref{sec:step2}, the column densities ($\log\,$\NH) in Table\,\ref{tab:cloudy_param} are not actual column densities in the simulations, but a parameter used to set the shape of the spectra. For each value of the \FUV flux, the cosmic ray ionization rate $\xi$ is scaled as
\begin{equation}
    {\rm \xi} =( F_{\mathrm FUV}/ G_0)   \xi_0 ,
\end{equation}
where $\xi_0=10^{-16}\,$s$^{-1}$ is taken as the average Milky Way value \citep{indriolo2007} and the solar neighborhood FUV flux $G_0=1.6\times10^{-3}$erg\,cm$^{-2}$\,s$^{-1}$ \citep{habing1968}. Exploring all combinations of the parameters listed in Table\,\ref{tab:cloudy_param}, leaves us with $129,600$ \cloudy one-zone models.

The one-zone \cloudy models must now be sampled according to the gas density PDFs found in Section\,\ref{sec:step3}, such that each combination of \nSFR and \nvw in the galaxy simulations corresponds to a density PDF-weighted sum of the 12 different one-zone \cloudy models along the $\log \nH$ axis that correspond to that region's other parameters ($\log Z$, $\log \NH$, $\log F_{\mathrm FUV}$ and $\log\,$DTM). For this step, six grid points in \nvw (from $10^{-4}$ to $10^2\,$\cc) and four grid points in \nSFR (from $10^{-2.5}$ to $10^{0.5}\,$\uSFR) are used. For each combination of \nvw, \nSFR we take the gas density PDF that comes closest in terms of both values, and shift the center of the lognormal to exactly match \nvw. This shift is performed to ensure that the PDFs generated are not only centered on the six chosen \nvw grid points, but can fill out the density space and generate a smooth total PDF as shown in Fig.\,\ref{fig:ex_PDFs}. The sampling of one-zone \cloudy models is carried out for all combinations of \NH, \FUV, $Z$, and DTM ratio, leading to a look-up table of $259,200$ different combinations of one-zone \cloudy models, with one such table for each spectral line considered. 

\begin{deluxetable}{l|ccc}
\tablecaption{Parameters for the One-zone \cloudy Models. \label{tab:cloudy_param}}
\tablehead{\colhead{Model Parameter}&\colhead{Min. value}&\colhead{Max. value}&\colhead{Step Size}}
\startdata
$\log\,$\nH [\cc]  & -4 & 7 & 1 \\
$\log\,Z$ [\Zsun]    & -2 & 0.5 & 0.5 \\
$\log\,$\NH [\cs]   & 17 & 22 & 0.5 \\
$\log\,$\FUV [\Go]  & -7 & 4 & 1 \\
$\log\,$DTM ratio  & -2 & -0.2 & 0.2
\enddata
\end{deluxetable}

\vspace{1cm}
\subsubsection{User-defined sub-grid attenuation function}\label{ext}
The \skirt calculation yields the average ISRF on scales similar to the resolution of the original cosmological simulation and hence parsec-size substructures are not taken into account by \skirt. 
As a way of compensating for this lack of resolution, we have included in the framework an optional user-defined function to add extinction on sub-grid scales. 
In practice, this is done by generating a one-zone \cloudy grid as described in Section\,\ref{sec:step4} but for which the incident spectra were attenuated using the ``extinguish'' command in \cloudy. 
This command diminishes the incident spectrum with a simple power law corresponding to an intervening slab of gas of fixed column density, which is set to 10$^{24}$\,\cs here. 

For the \simba simulations at hand, we test the framework with a very simple extinction function, which only adds extinction to \cloudy grid cells of density above $10^2$\,\cc. 
When constructing the look-up table by sampling the gas density PDFs (Section\,\ref{sec:step3}), these extinguished models can now be sample for a certain range in density ($\xi$ not being modified). For instance, all one-zone models with densities above a certain threshold can be assigned the attenuated spectrum to account for unresolved dense and shielded substructures.
This procedure roughly mimics a scenario in which high-density gas regions are the most shielded from ionizing photons. 
In the following Sections we will compare the results with and without this additional function.

\section{Testing of the code}\label{sec:val}

\subsection{The \zsam~ simulated galaxy sample} \label{sec:sample}

We apply \sigame to a simulated galaxy sample from the \simba cosmological galaxy formation simulation \citep{dave2019}.  The \simba simulations are run using the meshless finite-mass hydrodynamics technique in the \gizmo {\it N}-body plus hydrodynamics code \citep{hopkins2015,hopkins2017}.  The main \simba run evolves 1024$^3$ gas elements and 1024$^3$ dark matter particles within a $100 h^{-1}$~Mpc volume (\simba-100) from $z=249\to 0$.  To improve the dynamic range and test resolution convergence, we also use a higher-resolution run with 512$^3$ gas elements and 512$^3$ dark matter particles within a $25h^{-1}$~Mpc volume (\simba-25).  

\simba includes a range of sub-grid models for galaxy formation physics, including \molH-based star formation, two-phase kinetic galactic winds, torque-limited and Bondi black hole accretion, three types of AGN feedback, and a sub-grid model to form and destroy dust during the simulation run \citep{li2018,Li2019}.  The galaxy properties in \simba have been compared to various observations across cosmic time and shown to provide reasonable agreement \citep[e.g.][]{Thomas2019,Appleby2020,lower2020,montero2019}. For more details we refer to \cite{dave2019}. 

We extracted samples of galaxies from the $z=0$ snapshots of \simba-100 and \simba-25 using the \caesar galaxy and halo catalog generator, which identifies galaxies using a six-dimensional friends-of-friends algorithm applied to dense gas ($n>0.13$~cm$^{-3}$) and stars \citep{thompson2015}. Before making any selection cuts, a total of 49,215 and 2463 galaxies were found by \caesar for \simba-100 and \simba-25, respectively.  To ensure we have a reasonably well-resolved gas distribution within the galaxy, we only consider galaxies with $>300$ gas elements in both simulation boxes, corresponding to a gas masses of at least $5.6\times 10^9\,M_\odot$ and $0.7\times 10^9\,M_\odot$ in \simba-100 and \simba-25, respectively. 
For comparison the mass resolution of the \arepo-M51 simulation, which will be used to fragment the gas as described in Section\,\ref{sec:step3}, is a few solar masses. 
From these we select a test sample of \nsam galaxies from \simba-100 and another sample of \nsamm galaxies from \simba-25. The samples were selected to span a wide range in stellar mass, SFR, and gas mass.  Only galaxies found to have nonzero mean SFR over the past 100~Myr were included.

\begin{figure}[htbp]
\centering
\includegraphics[width=1.1\columnwidth]{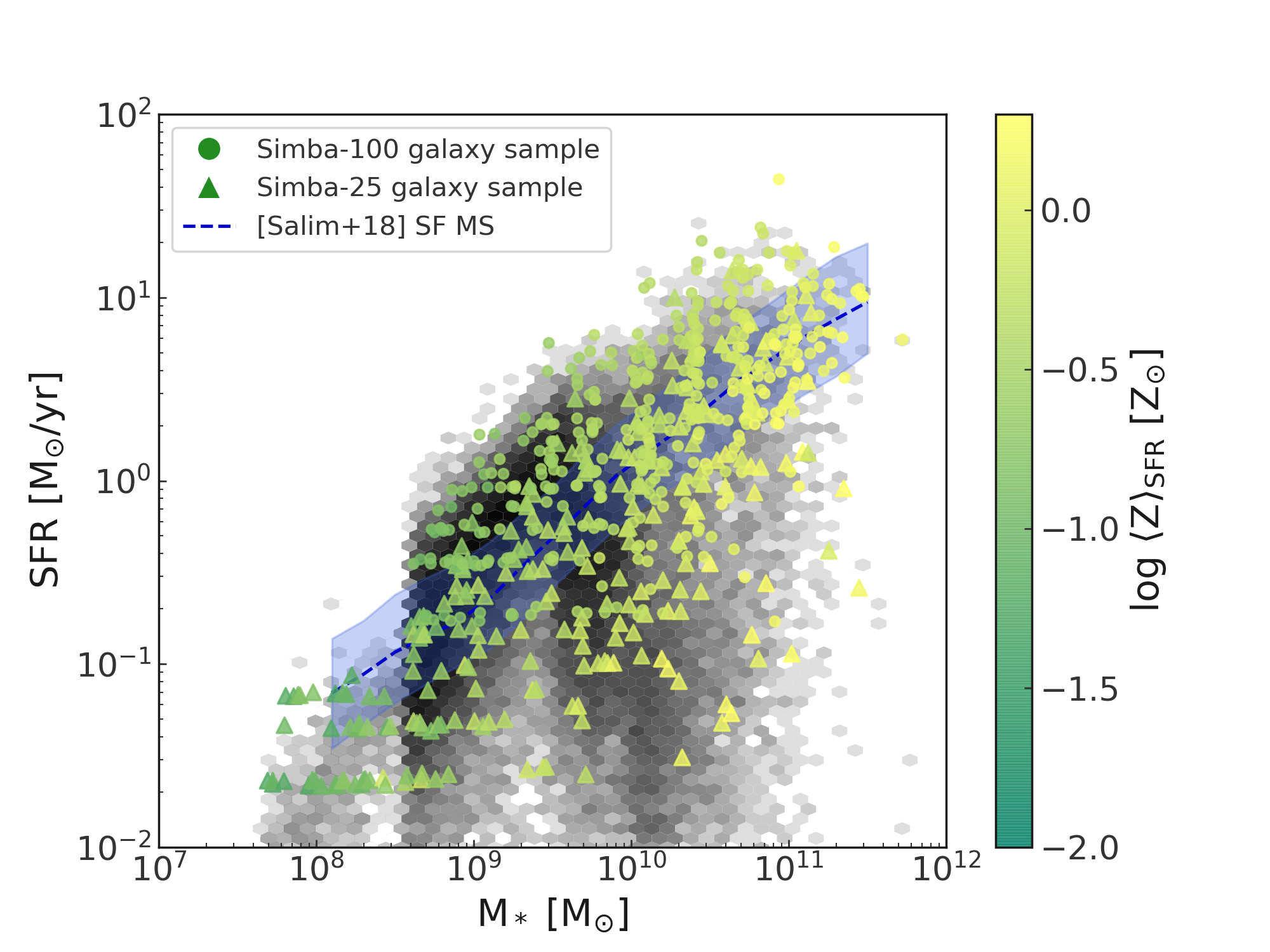}
\caption{The position on the SFR--\Mstar plane of the selected test sample at $z = 0$ of \nsam~\simba-100 galaxies (circles) and \nsamm~\simba-25 galaxies (triangles), with both samples color-coded by SFR-weighted metallicity. 
The stellar masses were calculated by summing over all stars within the six-dimensional structures generated by the friends-of-friends algorithm in \caesar. 
{For comparison, the gray hexbin contours show the overall distribution of $z=0$ star-forming galaxies in the combined volumes of \simba-100 and \simba-25, on a logarithmic grayscale.} The observed $z=0$ main sequence \citep{Salim2018} is also shown with a shaded region referring to 16th and 84th percentiles. \label{fig:Mstar_SFR}}
\end{figure}

\begin{figure*}[htbp]
\centering
\includegraphics[width=1.0\textwidth]{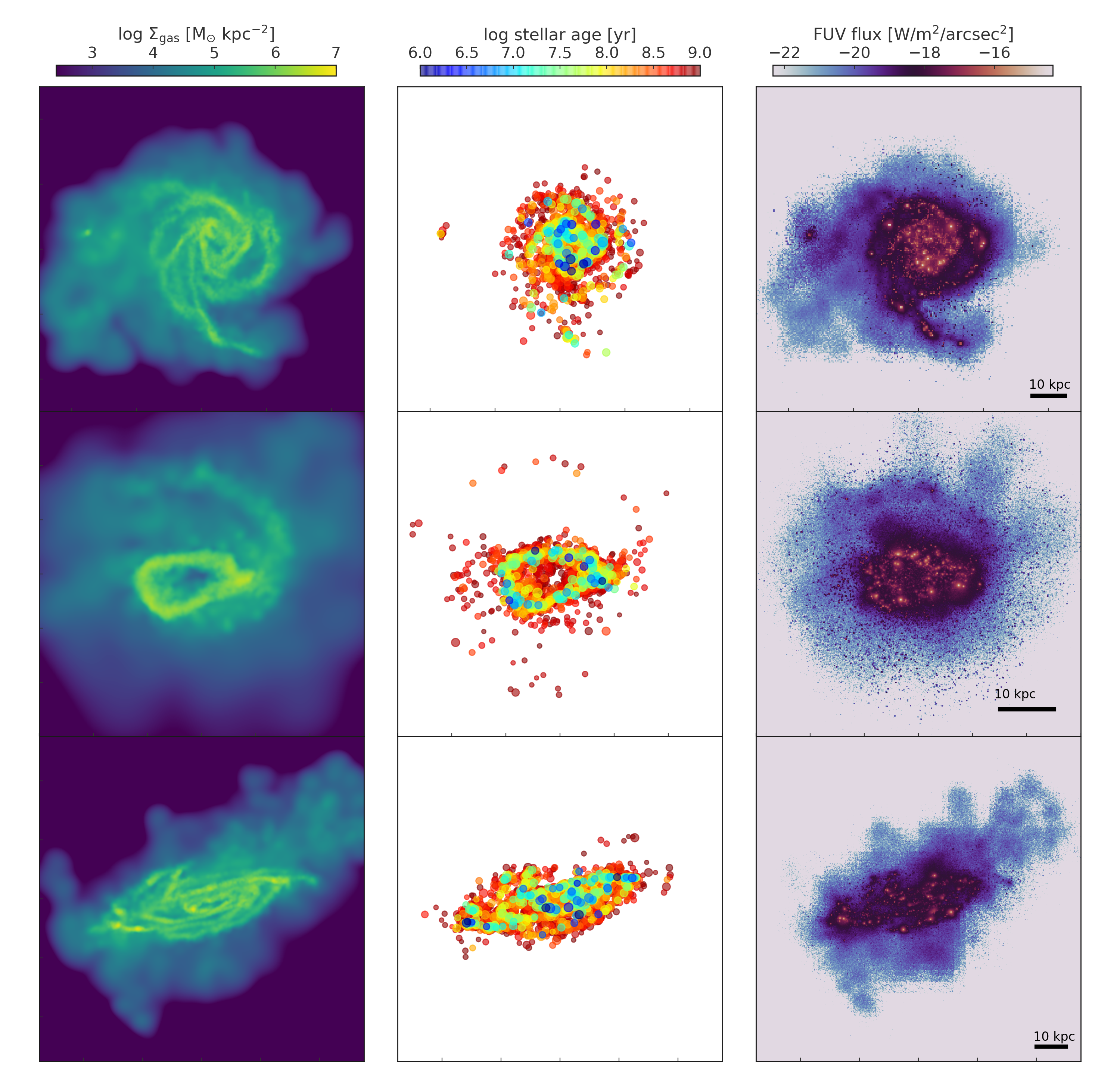}
\caption{Examples of three \simba-25 galaxies going through step 1 of \sigame, where stellar light is propagated, attenuated, and re-emitted by dust grains using the radiative transfer code \skirt. Left column: surface density maps of the total gas mass in \simba. Center column: positions and ages of all stellar population particles with ages below 1\,Gyr. The sizes of the stellar circles scale with stellar population mass. Right column: maps of observed FUV (6--13.6\,eV) flux derived with \skirt as observed from a distance of 10\,Mpc, showing increased FUV flux in areas close to young stellar populations and attenuation by regions of dense gas. \label{fig:maps}}
\end{figure*}

Figure\,\ref{fig:Mstar_SFR} shows the positions of the \simba-100 and \simba-25 galaxy samples in the SFR--\Mstar space.  The distribution of all \simba-100 and \simba-25 $z=0$ galaxies identified with \caesar and within the axis limits is shown underneath with logarithmic hexbin contours. 
The agreement with observations by \cite{Salim2018} is generally good, though \simba  overestimates the SFR over the range $\Mstar\sim10^{9.5}$--$10^{10}\Msun$~\citep[see discussion of this in][]{dave2019}. Overall, \simba reproduces the observed SFR--\Mstar distribution fairly well, both in terms of the main sequence and the quenched fractions~\citep{dave2019}, as well as the bimodality in specific SFR \citep{Katsianis2020}.

For our \simba-100 sample, hydrogen densities range from 0.5 to 66~\cc and SFR-weighted mean metallicities \Zsfr\ span 0.1--1.9\,\Zsun. By weighing the metallicity by SFR, we are approximating the metallicity that would be observed using nebular emission lines since these lines primarily trace star-forming regions. Global galaxy properties such as stellar mass, gas mass, SFR, \Zsfr, and radius can be found in Table\,\ref{tab:sample1} together with their line luminosities in Table\,\ref{tab:sample2} in Appendix\,\ref{app:100} and online. 

\begin{figure}[htbp]
\centering
\includegraphics[width=.5\textwidth]{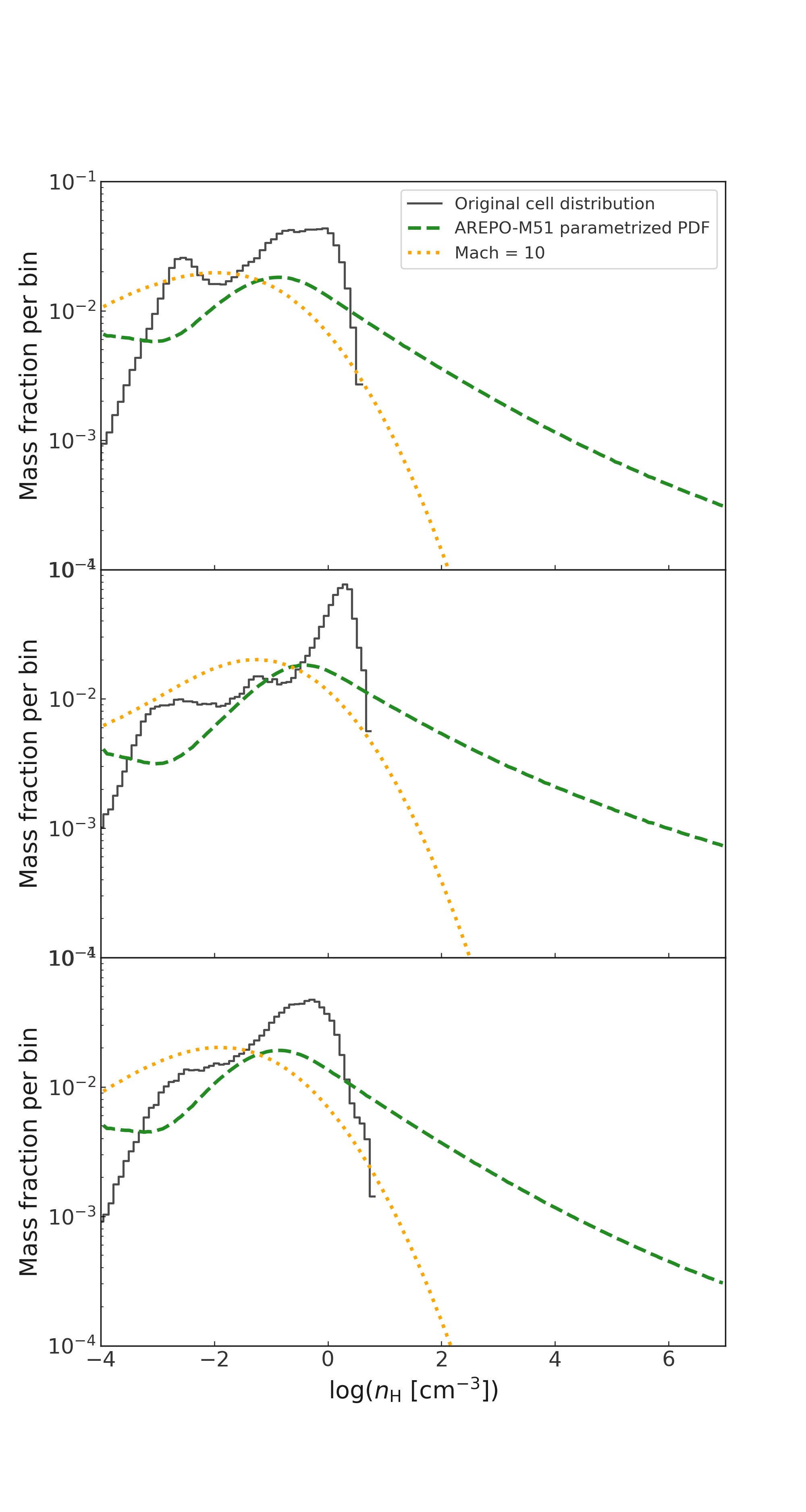}
\caption{Examples of density PDFs for the same three \simba-25 galaxies as shown in Figure\,\ref{fig:maps}, in the same order. Each panel compares the density PDF of the simulation, after regridding the particle data to a grid format with \skirt (solid histogram), the density PDF using the parameterized PDF from \arepo-M51 (dashed line), and the density PDF when adopting a lognormal of Mach number 10 for each gas cell (dotted line). \label{fig:ex_PDFs}}
\end{figure}

Figure\,\ref{fig:maps} illustrates the outcome of modeling and propagating the stellar light with \skirt in step 1 of \sigame (see Section\,\ref{sec:step1}). 
\skirt outputs the radiation in cells with sizes ranging from $19$\,pc to $6.3$\,kpc, and a mass-weighted distribution in cell sizes that peaks at $\sim200$\,pc (for the \simba-100 galaxies).
For the purposes of this paper, we found that a total of $10^8$ photons per galaxy was enough to give stable results, corresponding to more than $\sim2,500$ photons per star particle. See Appendix\,\ref{app:skirt} for a test increasing the number of photon packets to $10^9$ that resulted in a negligible change in total FUV luminosity and FUV flux distribution. 
For the dust component (``MediumSystem'' in \skirt), we directly use the dust masses that are calculated on-the-fly in \simba \citep{Li2019}. 
The resulting total infrared (3--1100$\,\mu$m) luminosities span from $2.41\times10^7$\,\Lsun to $2.73\times10^{11}$\,\Lsun and FUV (6--13.6\,eV) luminosities from $5.78\times10^7$\,\Lsun to $1.09\times10^{10}$\,\Lsun using the \simba-100 dust masses directly. 
The mass-weighted metallicities of our model galaxies range from 0.02 to 1.9\,\Zsun for the \simba-100 sample, so we set the metallicity of intervening gas to 1\,\Zsun in step 2 of \sigame (see Section\,\ref{sec:step2}).




After step 3 of \sigame (see Section\,\ref{sec:step3}), we can construct galaxy-wide density PDFs, and examples for the same galaxies as in Figure\,\ref{fig:maps} are shown in Figure\,\ref{fig:ex_PDFs}. 
The original density PDF in the cell data from \skirt (solid line) is compared to the derived PDF adopting a single lognormal with Mach number 10 and forcing parameter $b = 1/3$ versus the approach described in Section\,\ref{sec:step3} in dotted and dashed lines, respectively.

\begin{figure*}[htbp]
\centering
\includegraphics[width=1.\textwidth]{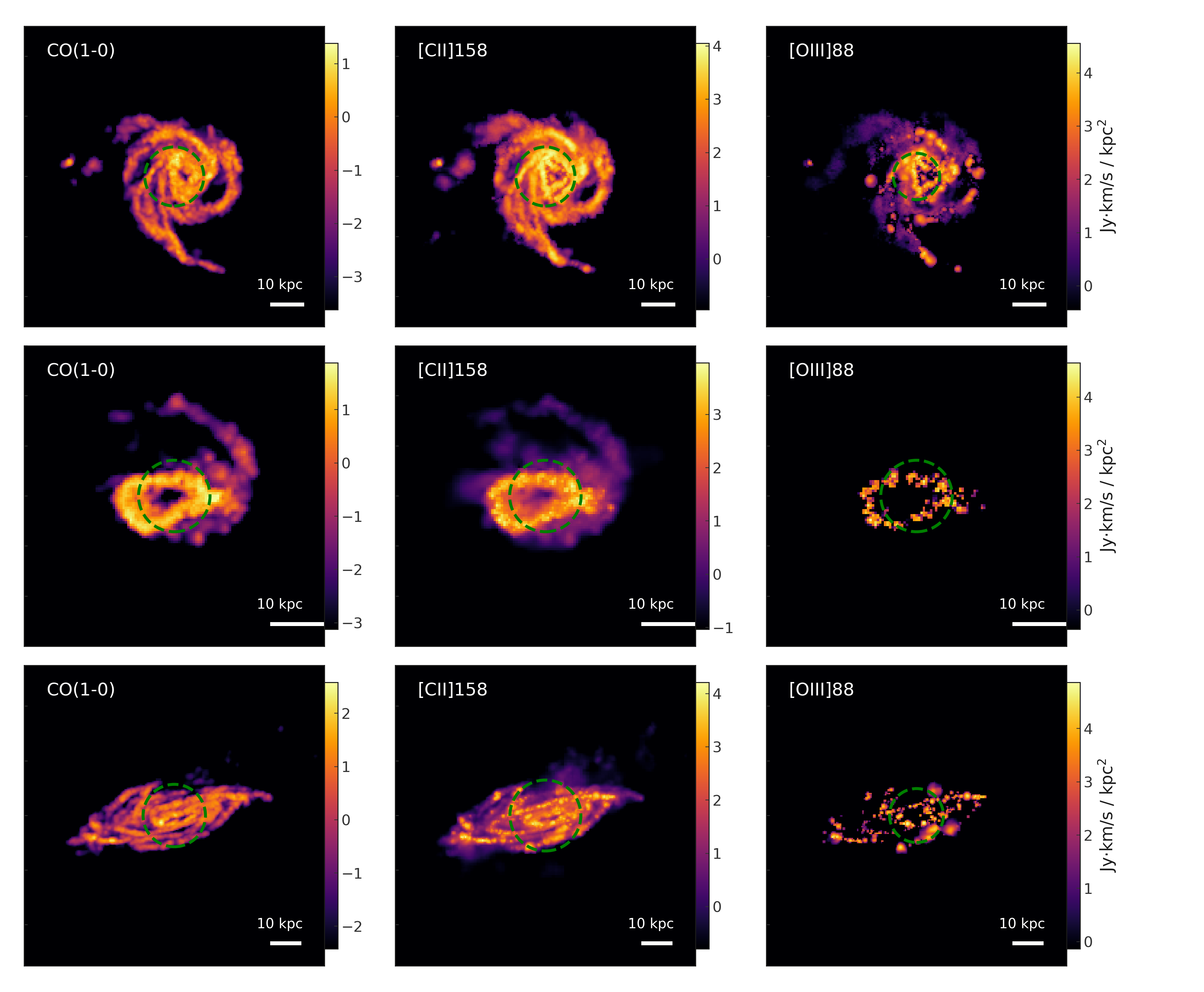}
\caption{Moment0 maps of the same galaxies shown in Fig\,\ref{fig:maps}, for three of the emission lines investigated here. Green dashed circles indicate the half-light radius. \label{fig:mom0}}
\end{figure*}

Fig\,\ref{fig:mom0} shows moment0 maps of CO(1-0), \cii and \oiii emission constructed for the same three galaxies from \simba-100 shown in Figure\,\ref{fig:maps}, using the output from completing step 4 of \sigame (see Section\,\ref{sec:step4}).
The output data cube from \sigame contains spatial information, cell sizes, and cell luminosities that were combined to derive the surface brightness of each pixel in the final image, weighting each cell by its volume filling factor within the column covered by each pixel\footnote{\url{https://github.com/jaymotka/moment0_maps}}. 
We note that the calculation of the moment0 maps assumes an ISM fully transparent to the FIR line emission, an assumption that might not hold for shorter wavelengths. Some expected differences in how the emission lines trace the ISM can be observed, such as the preference for \cii to arrive from PDRs near star forming regions compared to the broader CO(1-0) emission, and the relatively concentrated \oiii emission restricted to \hii regions experiencing harder radiation fields from young stars. 
The more concentrated \cii and \oiii emission relative to the CO(1-0) is also reflected in slightly smaller half-light radii compared to the CO(1-0) maps as illustrated with green dashed circles.

\subsection{\sigame runs}\label{sec:runs}

\begin{deluxetable}{l|p{4.5cm}}
\tablecaption{The Different \sigame Runs Compared in This Study.\label{tab:runs}}
\tablehead{\colhead{Run name}&\colhead{Description}}
\startdata
100Mpc\_M10         &  Adopting one lognormal with Mach number 10 with a forcing parameter 1/3 for the sub-grid density profile (see Section\,\ref{sec:step3}). \\
100Mpc\_SIGAMEv2    &  Comparison run with \sigame v2 applied to the \simba-100 sample. \\
100Mpc\_arepoPDF    &  Adopting density PDFs with a power-law tail as parameterized for a higher-resolution \arepo-M51 simulation (see Section\,\ref{sec:step3}). This is our ``default run''. \\
25Mpc\_arepoPDF       &  Same as ``100Mpc\_arepoPDF'' but for the \simba-25 simulation volume (see Section\,\ref{sec:sample}). \\
100Mpc\_arepoPDF\_no\_ext    &  Same as ``100Mpc\_arepoPDF'' but without the additional extinction function described in Section\,\ref{ext}. 
\enddata
\end{deluxetable}

In order to compare different assumptions in the sub-grid process, we perform four different runs on the same $z\sim0$ sample from the \simba-100 box as well as one run on galaxies selected in the \simba-25 box. The names of the different runs are listed in Table\,\ref{tab:runs}. 
In the first run (\ncode{100Mpc\_M10} in Table\,\ref{tab:runs}), we report the line luminosities that \sigame returns when the density fragmentation is carried out the ``traditional way'' with a single lognormal density PDF corresponding to a Mach number 10 and a forcing parameter of 1/3. 
A value of 10 for the Mach number is supported by observations of clouds in the solar neighborhood with typical sound speeds of 0.2--0.3\,km\,s$^{-1}$ and velocity dispersions of several km s$^{-1}$ \citep[see, e.g.][]{goldreich1974,kainulainen2013,hennebelle2019}. 
Next, we applied \sigame v2 to the same set of galaxies, using the modifications to the code described in \citet[][]{Leung2020} together with a new \cloudy grid at $z=0$ (\ncode{100Mpc\_SIGAMEv2}). 
The next simulation run adopts the new gas fragmentation scheme described in this paper, and will be referred to as our ``default run'' (\ncode{100Mpc\_arepoPDF}). 
The default settings are also applied to a smaller sample of \nsamm galaxies in the \simba-25 box as a convergence test (\ncode{25Mpc\_arepoPDF}). 
Finally, we also investigate the resulting line emission when not including the additional sub-grid attenuation function as described in Section\,\ref{ext} and otherwise adopted in the default run (\ncode{100Mpc\_arepoPDF\_no\_ext}).

\subsection{Observed galaxy sample for comparison}\label{sec:obs}
We will be comparing the different simulation runs with a broad selection of nearby observed galaxies. 
Since the samples of simulated galaxies were selected to span a wide range in stellar mass, SFR, and gas mass, they were not tailored to match any specific observed sample. 
We will be comparing the range in simulated and observed line luminosity as a function of SFR, looking for agreements on the order of magnitude and deferring a more careful comparison with observations to a future study. 
The sample of observed galaxies comprises all nearby galaxies observed with Herschel or the Infrared Space Observatory that we could find in the literature for which SFR estimates could also be made. In order to best compare with observations, we estimate the SFR of the model galaxies as a mean over the past 100 Myr, and not the instantaneous SFR of gas particles. Furthermore, all SFRs derived using a \citet{Salpeter1955} IMF have been converted to the equivalent SFR of a \citet{Chabrier2003} IMF as in the \simba simulation, by adopting a $-0.24$\,dex correction \citep[e.g.][]{mitchell2013}. 
The observed SFRs were obtained with various methods summarized here; 
For the sample of mixed-type galaxies in \cite{Kamenetzky2016} and the uminous infrared galaxies (LIRGS) in \cite{Diaz-Santos2013}, the FIR luminosities, \Lfir ($40$--$120\,\micron$ and $42.5$--$122.5\,\micron$ in the two respective cases), of those papers are converted into SFRs according to \cite{Kennicutt1998}. 
For the sample of dwarf galaxies of \cite{Cormier2015}, SFRs come from \cite{remy2015}, who derived SFR from FUV and H$\alpha$ luminosities, corrected for dust attenuation. 
The galaxy sample of \cite{schruba2012} contains CO(2-1) luminosities for 16 dwarf galaxies (with one galaxy in common with the \cite{Cormier2015} sample) as well as CO(1-0) luminosities for nearby galaxies compiled from the literature, of which we show 22 CO(1-0) observations from the HERA CO-Line Extragalactic Survey \citep[HERACLES;][]{leroy2009} and 20 from \cite{calzetti2010}. 
For the 16 dwarf galaxies, SFRs in \cite{schruba2012} were calculated as a combination of FUV and $24\,\mu$m
emission following the approach in \cite{bigiel2008} and \cite{leroy2008}. 
For the SFRs of the 50 nearby galaxies we use the values compiled from literature in \cite{schruba2012}, and refer to the references therein for specific methods. 
Finally, for the sample of main-sequence and starburst galaxies of \cite{Brauher2008}, we convert the reported 63\,$\micron$ flux to an SFR using the 70\,$\micron$ monochromatic SFR conversion relation of \cite{calzetti2010}.

\subsection{Comparison of simulation runs and observations}\label{sec:comp}
\begin{figure*}[htbp]
\centering
\includegraphics[width=1\textwidth]{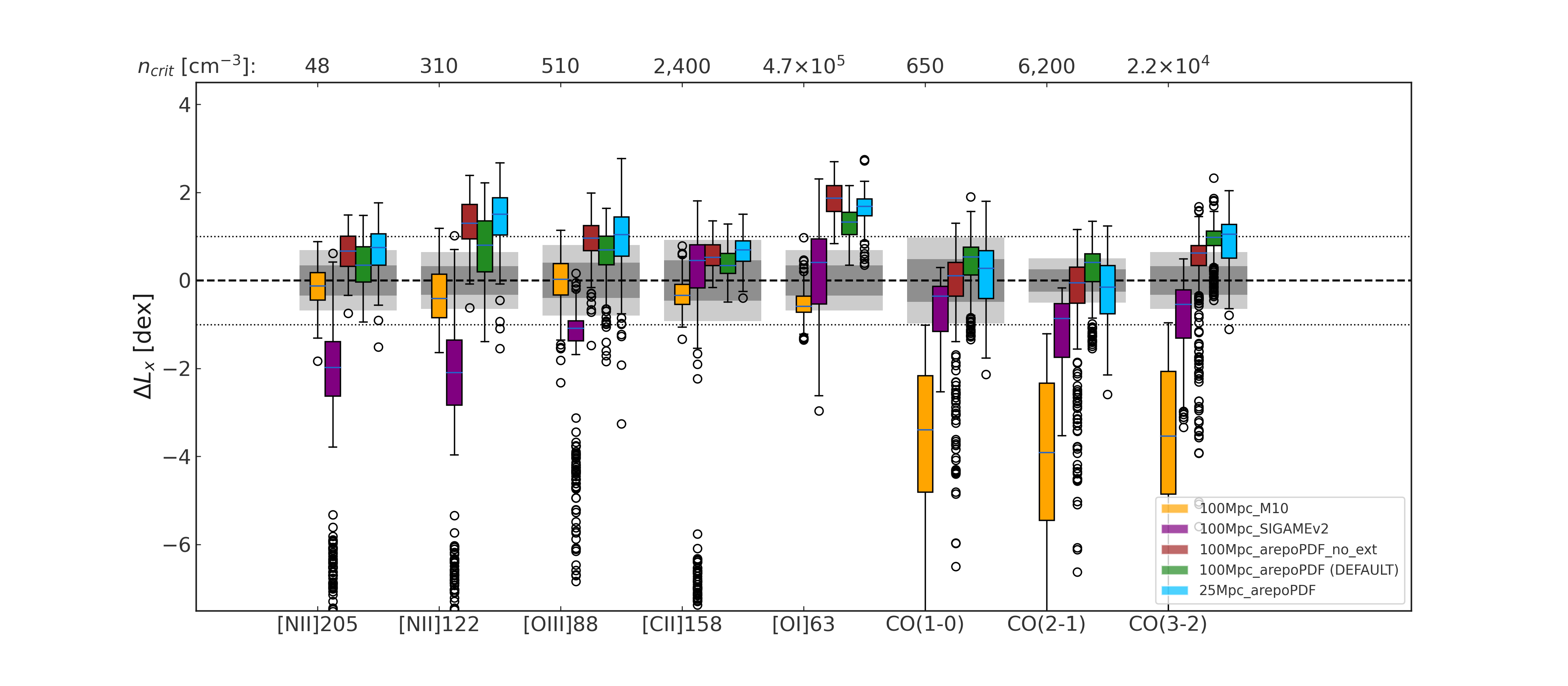}
\caption{Box plot performance comparison of the different \sigame runs listed in Table\,\ref{tab:runs} for all lines considered here, with the fine-structure lines sorted in order of increasing critical density. Light and dark gray shaded areas represent 1$\sigma$ and 3$\sigma$ spread in the observations. The boxes extend from the lower to the upper quartile values, whiskers to a factor $1.5$ wider range, and open circles indicate outliers outside the whiskers. Horizontal lines within each box indicate the median value, and the dotted horizontal lines show +/- 1\,dex. The critical densities are from Table\,\ref{tab:crit} with $\rho_{\mathrm crit}$ of atomic hydrogen for \cii. 
The number of observed galaxies used for each of emission lines varies from $21$ to $455$ and they span galaxies of different types: LIRGs \citep{Diaz-Santos2017}, dwarf galaxies \citep{Madden2013,Cormier2015}, main-sequence and starburst galaxies \citep{Brauher2008}, (ultra)LIRGs \citep{Farrah2013,Zhao2016}, the randomly selected $0.01<z<0.02$ COLD GASS galaxy sample of \citet{accurso2017}, and the mixed sample of \citet{Kamenetzky2016} and \cite{schruba2012}. For the dwarf galaxy sample, we only include seven galaxies with galaxy-integrated luminosities, following the criteria of \cite{accurso2017}. For the sample of \cite{Kamenetzky2016}, we include only galaxies with optical sizes smaller than 47'' as a very conservative measure to ensure that the {\it Herschel}/PACS field of view included all line emission. See the text for how SFR was calculated. \label{fig:std}}
\end{figure*}

The comparison between simulation runs relative to observed line luminosity--SFR relations is made in Figure\,\ref{fig:std}. 
The figure shows the offset in line luminosity from a power law fit made to the observed line luminosity as a function of SFR for each of the eight emission lines considered here. 
The standard deviation of the offsets for the observed galaxies is indicated with gray bars in the background. 
The different \sigame runs listed in Table\,\ref{tab:runs} can now be compared for each of the eight emission lines considered here, with the fine-structure lines sorted in order of increasing critical density (Table\,\ref{tab:crit}). 
Due to the mixed sample of observed galaxies that we compare to, we define a ``good agreement'' here as when the quartile distribution of simulated line luminosities overlaps with the 2-$\sigma$ spread around the observed line luminosity--SFR relation.

For run \ncode{100Mpc\_M10}, the resulting deviations from observations are shown with orange bars. Surprisingly, this simple assumption does a very good job for the ionized species of \niiA, \niiB, \oiii and \cii, for which a good agreement is reached, but it completely underestimates CO emission. The CO(1-0), (2-1) and (3-2) lines fall below the observed line-SFR relation by median deviations of $3.4$, $3.9$ and $3.5$\,dex, respectively. 
The underestimation of the molecular lines is not surprising, given that the density PDF does not extend significantly beyond densities $>10^2$\,\cc where molecules form (see Fig.\,\ref{fig:ex_PDFs}). 

With run \ncode{100Mpc\_SIGAMEv2} (purple bars), we test how well the previous \sigame framework, which has so far only been applied at $z\sim2$ and above, works for the \simba-100 $z=0$ sample. While this version of the code gives a good agreement with observations for \cii, \oi, CO(1-0) and CO(3-2), it significantly underestimates the \nii~and \oiii lines with median deviations larger than 1\,dex. For all lines except \oi, this version also produces outliers with extremely low line luminosities, as shown with open circles (and indeed, the plot does not extend to show all outliers). We attribute this disagreement to the simplified gas fragmentation scheme used in \sigame v2 in which gas particles were split into either GMCs following a locally observed mass spectrum or diffuse gas, with no intermediate ISM phase and no sanity checks on the density PDF.

The default run of the current version of \sigame, \ncode{100Mpc\_arepoPDF} (green bars) comes closest to all line luminosities simultaneously, with the fewest outliers. 
In total, six out of the eight lines considered here are in good agreement with the observed relations, while the lines of \oi and CO(3-2) line luminosities are overestimated by median deviations of $1.3$ and $1.0$\,dex, respectively, compared to the observations. 
We do not consider the overestimation of CO(3-2) an outstanding problem due to the lack of observations compiled here (our observed comparison sample comprises 21 galaxies). 

\begin{figure*}[htbp]
\centering
\includegraphics[width=1\textwidth]{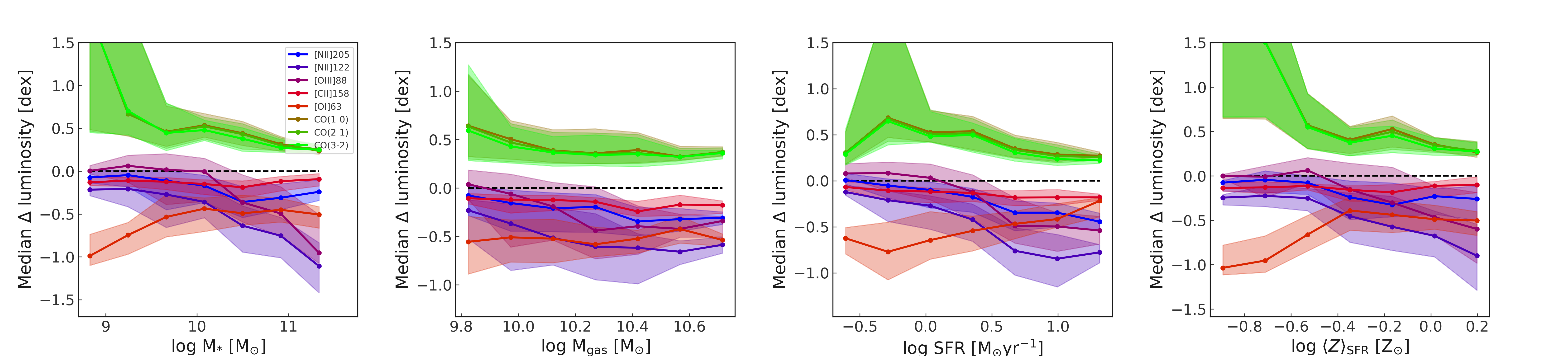}
\caption{Line luminosities in run \ncode{100Mpc\_arepoPDF\_no\_ext} subtracted from those in run \ncode{100Mpc\_arepoPDF} (our default run) as a function of galaxy properties \Mstar, \Mgas, SFR, and \Zsfr. Solid lines give the median deviation in luminosity whereas shaded regions show the range in luminosities from the 25th to 75th percentile. The CO line luminosities of \ncode{100Mpc\_arepoPDF} are consistently above those of \ncode{100Mpc\_arepoPDF\_no\_ext}, and vice versa for the rest of the (non-molecular) emission lines. \label{fig:ext1}}
\end{figure*}

The test run without additional extinction, \ncode{100Mpc\_arepoPDF\_no\_ext} (brown bars), looks very similar to the default run in Fig.\,\ref{fig:std} with the most noticeable difference being an increased \oi line prediction that falls above the observed luminosities by a median deviation of 1.9\,dex. 
In our default run, with the user-defined attenuation function described in Section\,\ref{ext}, the overestimation of \oi is brought down by about $0.5$\,dex to $1.3$\,dex above the observed \oi--SFR relation, and the tail of low CO luminosities is reduced in length by several dex. This suggests that the inclusion of additional attenuation is necessary in order for enough \ion{O}{1} and CO to form. 
An effect on the \nii and \oiii lines could also be expected since these lines originate in dense \hii regions. Although the current version of \sigame does not explicitly model \hii regions (see the discussion in Section\,\ref{hii_regions}), the additional attenuation at densities above $10^2\,\cc$ should also affect these lines, but this is not clear from the distributions in Fig.\,\ref{fig:std} alone. 
In order to investigate the effect of additional attenuation, Fig.\,\ref{fig:ext1} compares the default run (\ncode{100Mpc\_arepoPDF}) to that without additional attenuation (\ncode{100Mpc\_arepoPDF\_no\_ext}) as a function of galaxy properties \Mstar, \Mgas, SFR, and \Zsfr. 
Only the CO and \oi line luminosities are consistently higher and lower, respectively, in the default run, in agreement with Fig.\,\ref{fig:std}. 
The deviations in the \nii, \oiii and \cii line luminosities are negligible at the lower end of \Mstar, \Mgas, SFR and \Zsfr, but tend toward negative values at the higher end. 
In particular, the \nii and \oiii lines are more strongly affected by the additional attenuation for high values of SFR and \Zsfr (rightmost two panels). 
This behavior can be explained by considering what happens in the one-zone \cloudy models of which our look-up table consists. 
At higher metallicities, increasing self-shielding of nitrogen and oxygen results in more singly ionized nitrogen and doubly ionized oxygen relative to higher ionization states since the one-zone \cloudy models include self-shielding. 
This in turn results in higher \nii and \oiii luminosities with increasing metallicity for a given one-zone \cloudy model and hence a strong dependence on metallicity in run \ncode{100Mpc\_arepoPDF\_no\_ext}. 
However, once additional attenuation is added at densities above 100\,\cc, the luminosity of those one-zone models is negligible as there is no ionizing radiation to ionize nitrogen and oxygen. 
The difference between the two runs is therefore more pronounced at high metallicity, where models with densities above 100\,\cc have had their \nii and \oiii luminosities dramatically reduced. Since there is a weak correlation between SFR and \Zsfr in our sample, the same can be said for the evolution with SFR seen in Fig.\,\ref{fig:ext1}.



Finally, the test run with default settings at higher mass resolution, \ncode{25Mpc\_arepoPDF} (blue bars), shows quartile ranges that overlap with those of the \simba-100 sample, indicating that the \sigame results do not depend significantly on mass resolution in the underlying simulation. 
However, this analysis alone does not reveal any bias due to the different galaxy properties of the two samples. In Fig.\,\ref{fig:25-100} in Appendix\,\ref{app:25-100} we investigate in more depth how well the two simulation runs agree with one another for a range of physical galaxy properties, finding that for similar galaxies, the luminosities of \niiA, \niiB, \oiii, \cii and \oi tend to be overestimated by up to $\sim0.5$--$1.5$\,dex in \simba-25 compared to \simba-100, in particular at high SFR and \Zsfr. 
It is interesting to note that \ncode{100Mpc\_arepoPDF\_no\_ext} behaves more similarly to \ncode{25Mpc\_arepoPDF} than to \ncode{100Mpc\_arepoPDF}, and it is hard to say whether this is a resolution issue or due to the attenuation function present in the latter two runs.

\section{Discussion} \label{sec:dis}


\subsection{Comparison with previous models}\label{sec:comp_models}
\begin{figure}[htbp]
\centering
\includegraphics[width=1\columnwidth]{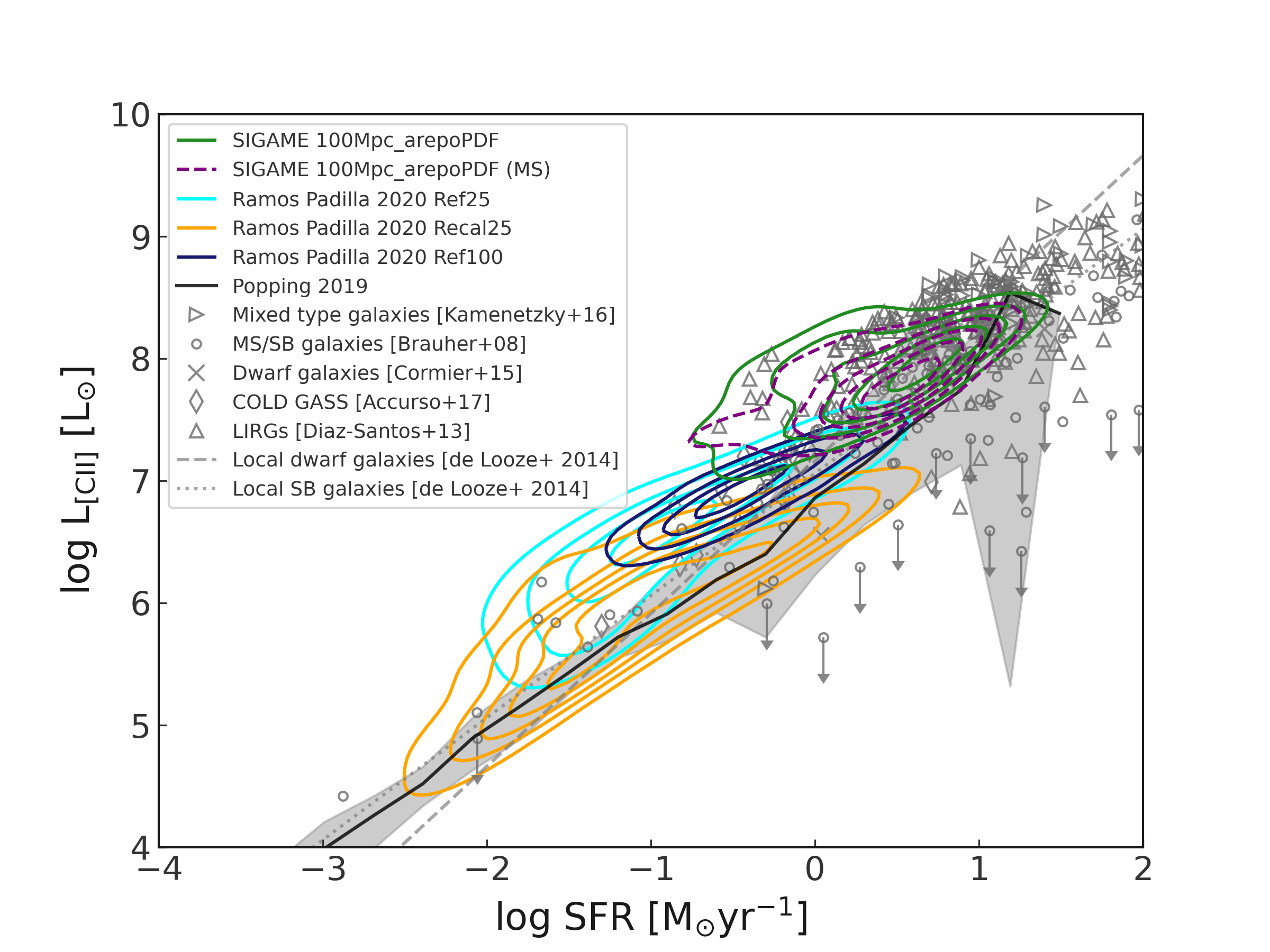}
\caption{Comparison of the \sigame v3 \cii--SFR relation (\ncode{100Mpc\_arepoPDF}: green; MS-only galaxies: purple dashed contour lines) with other recent modeling efforts: a multiphase model of the ISM used as a postprocessing step for the EAGLE cosmological hydrodynamical simulations \citep[cyan, orange, and dark blue contour lines;][]{ramospadilla2020}, and the $z=0$ \cii--SFR relation found by \citet{Popping2019} using semianalytical models (black line), with a 1$\sigma$ shaded region. Observations of nearby galaxies are shown with gray symbols \citep{Brauher2008,Diaz-Santos2013,Cormier2015,Kamenetzky2016,accurso2017} and the \cii--SFR relations for local starburst and dwarf galaxies by \citet{delooze2014} are shown with gray dotted and dashed lines, respectively.
\label{fig:models}}
\end{figure}

We can compare our \cii--SFR relation with that found through similar techniques and hence better understand the decisive differences between the models. 
Figure\,\ref{fig:models} shows the \cii--SFR relations found by \sigame for the \simba-100 galaxies with the \ncode{100Mpc\_arepoPDF} and run (green contour lines) together with the relations derived by \cite{Popping2019} and \cite{ramospadilla2020}. 
From the \ncode{100Mpc\_arepoPDF} run, we also show the distribution of only main-sequence (MS) galaxies, chosen to be between the 16th and 84th percentile of the \cite{Salim2018} relation (purple dashed contour lines). 
The distribution of \sigame simulated galaxies generally matches that of the observed galaxies well, but fails to reproduce the some of the galaxies of lower \cii luminosity and the nondetections of \cite{Brauher2008}. 
Selecting only MS galaxies does not change this picture significantly. 
Our simulated galaxies lie in above the simulated sample of \cite{Popping2019}, but in apparent extension of the samples of \cite{ramospadilla2020}, made with cosmological simulations from the Evolution and Assembly of Galaxies and their Environments (EAGLE) project \citep{crain2015,schaye2015}. In the latter study, GMCs are created in postprocessing following \cite{Olsen15a} and \cloudy is used to derive line intensities as in newer versions of \sigame. 
One interesting avenue of further study would be to apply \sigame to EAGLE, to fully understand how intrinsic differences between EAGLE and \simba affect this comparison.

Another interesting comparison can be made to the CO line modeling of \cite{Vallini2018} (hereafter V18) in which the sub-grid density PDF was determined using the Mach number of the underlying simulation together with a time-dependent evolution of the high-density power law tail. 
V18 find good agreement with the observed relation between GMC virial mass and CO luminosity in nearby galaxies, the CO excitation in nearby starburst galaxies and the CO excitation of a low-metallicity GMC in the LMC. 
In contrast, our \ncode{100Mpc\_M10} run, which comes closest to the method of V18 does not reproduce the CO line emission well. 
Two key differences that likely cause this difference are: (1) in V18, the Mach number is allowed to change and is in fact inherited from the underlying simulation, and (2) the underlying simulation used in V18 is very different from \simba in the sense that they reach much higher densities before any postprocessing is performed. The density PDF of Alth{\ae}a (the \ramses zoom-in simulation used by V18) peaks at 300\,\cc and reaches densities above 1000\,\cc. In our case, \simba hardly reaches 100\,\cc, and hence using a lognormal of any Mach number (even one higher than 10) does not result in the densities high enough to excite CO. It could be said that the approach in V18 works well if the underlying simulation reaches certain minimum densities, but for cosmological simulations, applying a lognormal (+powerlaw) centered at the mean volume-averaged cell density will not be enough on its own.

\subsection{Choice of simulation for the fragmentation task}\label{sec:CMZ}
We have only shown results for the present framework using one type of simulation, namely the \arepo suite of M51 realizations. We also examined an \arepo simulation of the central molecular zone (CMZ), the \arepo-CMZ simulation \citep{sormani2020}. The \arepo-CMZ simulation contains relatively compact clouds with extreme velocity dispersions and a threshold density for the formation of sink particles an order of magnitude higher than that for the M51 simulation, meaning that the ISM is allowed to fragment further before forming sink particles. By binning the density PDFs of the \arepo-CMZ simulation in the same way as done for the M51 simulation (Figure\,\ref{fig:AREPO_PDFs}), we can compare the PDF shapes and the resulting line luminosity for two different simulations. At high volume-averaged gas densities, the mean PDF shapes are similar between the two simulations, but at lower densities ($\log\nH < 0$), the \arepo-CMZ simulation returns PDFs with relatively more mass at higher densities. The resulting line luminosities using the \arepo-CMZ PDF table are in general higher (up to $0.27$\,dex on average for \oiii) except for the CO lines which are lower by up to $0.19$\,dex on average. In Fig.\,\ref{fig:line_nH} in Appendix\,\ref{app:line_nH} we show the line luminosity for bins in mass-weighted hydrogen density for all cells in the \simba-25 galaxy shown in the top panels of Fig.\,\ref{fig:maps} with different assumptions for the sub-grid density PDF. We conclude that the galaxy region sampled can be important, although a better test would be to compare with an entirely different galaxy simulation (not just a special region like the CMZ). 

Another caveat associated with the approximations used for the \arepo-M51 simulations is a tendency for the clouds to be too long-lived, which might in turn yield too little SFR for a given surface density. Of relevance to \sigame, this could mean that we are overestimating the number of clouds (i.e. dense gas mass fraction in the PDF) for a given SFR volume density. 
However, choosing a completely different galaxy simulation technique \citep[e.g.\ models run with adaptive mesh refinement codes such as \ramses, \enzo, or \athena---][]{teyssier2002,bryan2014,kim2017} might return different PDFs, the effect of which we have not studied. Instead, we have focused on making the current version of \sigame flexible and publicly available so that the user can include any high-resolution simulation for the gas fragmentation if they so desire.

\subsection{Small-scale radiation field structure}\label{hii_regions}
As discussed in the previous section, this version of \sigame fragments the gas density on scales smaller than the \skirt cell sizes (typically larger than $\sim15$\,pc), thereby allowing for clumps of higher densities on scales not resolved by the parent simulation. 
In a similar manner, we could also expect the radiation field to be fragmented such that the mass-weighted FUV flux distribution on parsec scales can differ substantially from the overall cell mean flux value and allow for parsec-size features such as \hii regions or shielded, dense, molecular cores. 
Judging from the overall agreement between simulated and observed lines, we speculate that most of the FIR lines considered here arise from structures of larger extent where parsec resolution in radiation is not necessary, with the exception of \oi, which is overestimated. 
The continued overestimation of the \oi and CO(3-2) line emission by our default run (see Fig.\,\ref{fig:std}) suggests that a more careful treatment of the sub-grid attenuation and/or other features remains missing, however. 
A proper treatment of the flux distribution on these scales would require considerations about the properties of natal clouds around young stars, as demonstrated with the one-dimensional stellar feedback model \ncode{warpfield} \citep{Rahner2017}. 
In addition, \cii and \oiii remain slightly overestimated, both of which originate mainly in the atomic ISM phase in our model, hinting at an overestimation of atomic gas mass versus ionized gas mass in our gas fragmentation scheme.

\subsubsection{Missing treatment of active galactic nuclei}
While the coevolution of AGNs and their host galaxies is simulated in \simba, the current version of \sigame does not include any effects of AGN presence, such as additional heating and radiation. The additional X-ray heating has been shown to increase excitation of high-$J$ CO lines at high redshifts \citep{Vallini2019}.  

\subsection{Missing shock-heating of the gas}
In starburst galaxies and mergers, shocks are expected to act as an additional heating source in the ISM. Although the next version of \cloudy may include a treatment for shock-heated gas the current version of \cloudy iterates to find a thermal and ionization equilibrium, thereby ignoring any shocked state of the gas that might be present in the simulation. One way to treat shock-heated gas, would be to set up a separate grid of models, following the technique used for \mappings \citep{allen2008}. Turbulence (2--10 km\,s$^{-1}$ in velocity dispersion) and/or density-dependent magnetic fields are also believed to play a role in setting the \oi emission, through their effect on line width, shielding and pumping, and the chemical and thermal state of the gas \citep{Canning2016}. However, testing the effect of these would require \cloudy models of more than one zone where optical depth is taken into account. 

\section{Conclusions} \label{sec:con}

This paper introduces an improved algorithm for estimating FIR line emission from large-volume cosmological galaxy simulations using coarse numerical resolution.
We postprocess such simulations using a sub-grid model that estimates the distribution of dense gas up to densities of $\sim10^7\,\cc$. This fragmentation scheme results in gas at lower densities being compressed to higher densities, and hence affects emission lines tracing all ISM phases. We test the scheme on eight low-to-medium-density ISM tracers.

The density distribution on sub-grid scales is modeled by sorting and parameterizing resolved regions in higher-resolution simulations and interpolating on the parameterized functions to set the density distribution for the cosmological simulations. This statistical approach avoids any assumption about the size and shape of molecular clouds, the turbulence within those clouds, or the existence of pressure equilibrium between the dense and more diffuse ISM phases. As a demonstration of this scheme, we use data with subparsec resolution from a model of M51 with \arepo \citep{tress2020}. Density PDFs are sampled in 200\,pc regions and binned in terms of the volume-averaged hydrogen and SFR density of each region. The resulting mean PDFs are parameterized and used to generate PDFs by interpolation for each resolution element of the cosmological simulation to generate gas densities $\sim10^2$--$10^7$\,\cc otherwise not present in the cosmological simulation results. As a future extension, this approach could also be used to develop
a new SFR prescription for cosmological simulations.

The local radiation field strength is determined with the radiative transfer code \skirt, which calculates the local dust-attenuated stellar spectrum. The \swiftsimio package is used to map gas properties from the cosmological simulation output to the cell structure from \skirt. 
Additional attenuation by gas is implemented through a set of \cloudy models, by matching the transmitted spectrum in the OIR-to-FUV of a certain column density of gas to the dust-attenuated spectrum of \skirt for each gas cell. 
Line emission and chemical information are derived from an extensive grid of $129,600$ \cloudy one-zone models, sampled according to the local density PDF to create a library totalling $259,200$ models of different combinations of \nH, \nSFR, \NH, \FUV, $Z$ and DTM ratio. 
Finally, to compensate for lack of resolution in the cosmological simulations used here, we add attenuation on sub-grid scales with a simple function that can be modified by the user to work on other simulation types. In this case, we add attenuation for \cloudy grid models with $\nH > 10^2$\,\cc. 

We test the method on the \simba cosmological simulations by extracting galaxies that span a wide range in stellar mass, SFR, and $Z$. 
As a rough validation of the method, results are compared to observed relations between line luminosity and SFR for a diverse sample of nearby galaxies and eight different emission lines. 
To test how well the method converges for different mass resolutions, we apply \sigame to the \simba 100 and 25\,Mpc volumes, the latter of which has roughly eight times higher mass resolution for the gas particles. 
We also compare with results using the previous version of \sigame (v2) and results when not including the user-defined attenuation function. 
Finally, the method is also checked against the default option of adopting a lognormal density profile drawn from Mach 10 isothermal turbulence for the sub-grid density profile, although we expect this method to fare poorly in the case of cosmological simulations. These tests leave us with the following conclusions.

\begin{itemize}
\item The novel method presented here of using a high-resolution simulation and a new analytic PDF approach for the gas fragmentation results in emission lines that are in good agreement (i.e. the quartile values overlap with the $2\sigma$ spread around the observed line--SFR relation) for all but the \oi and CO(3-2) lines that are overestimated by on average $1.3$, $1.0$\,dex, respectively.
\item We find that the resulting line emission of \oi in particular is highly dependent on the user-defined attenuation function, without which the overestimation of the \oi luminosities is about 0.5\,dex higher still. 
This underlines the need for a more careful treatment of the radiation on sub-grid scales ($<200$\,pc), where denser regions should produce additional attenuation of the interstellar radiation field than what is included in this framework by default.
\item We find a good agreement with observations of nearby galaxies and other models in terms of \cii--SFR relation, although our default model cannot reproduce the non-detections of \cii.
\item Comparing line luminosities of galaxies in the \simba-100 and \simba-25 samples, we find that the latter returns higher luminosities by up to $\sim0.5$--$1.5$\,dex, in particular at high SFR and SFR-weighted $Z$.
\item Comparing with the previous version of \sigame (v2), we find that this method significantly underestimates the \nii and \oiii lines with median deviations larger than 1\,dex, most likely due to a simplified ISM structure.
\item The standard method in the literature of adopting the density PDF of Mach 10 isothermal turbulence for fragmenting gas on sub-grid scales results in FIR fine-structure line emission that agrees well with observations for lines originating mainly in the ionized ISM, but drastically underestimates lines from the neutral and molecular regions. For example, the CO(1-0), (2-1) and (3-2) rotational lines fall below the observed line-SFR relation by median deviations of $3.4$, $3.9$ and $3.5$\,dex, respectively.

\end{itemize}

We have presented \sigame v3 which is a flexible framework that can be adapted to any cosmological or galaxy simulation, and now has the option to further fragment the gas on sub-grid scales using a high-resolution simulation of choice. This tool may be useful for the interpretation of current data from e.g. ALMA, NOEMA and SOFIA as well as from future space and balloon missions such as JWST, GUSTO, and ASTHROS.

\acknowledgments
The authors thank the anonymous referee for thoughtful suggestions and comments that improved the paper. 
The authors thank Dr. Peter Camps for extensive support on \skirt, Robert Thompson for developing \caesar, and the \yt team for development and support of \yt \citep{Turk2011}. The authors are also grateful to the entire \simba collaboration who helped in the analysis and understanding of the \simba output. We acknowledge use of the \python programming language version 3.7, available at \url{http://www.python.org}, Astropy \citep{astropy:2013,astropy:2018}, Matplotlib \citep{Hunter2007}, NumPy \citep{harris2020}, pandas \citep{McKinney2010}, SciPy \citep{Virtanen2020SciPy-NMeth} and \swiftsimio \citep{Borrow2020}. 
This research has made use of the NASA/IPAC Extragalactic Database (NED), which is funded by the National Aeronautics and Space Administration and operated by the California Institute of Technology. 
This work made use of v2.2.1 of the Binary Population and Spectral Synthesis (BPASS) models as described in \cite{Eldridge2017} and \cite{Stanway2018}.
K.P.O. is funded by NASA under award No 80NSSC19K1651. 
B.B. is partly funded by the Packard Fellowship for Science and
Engineering. 
J.B. is supported by STFC studentship ST/R504725/1. 
R.J.S. acknowledges an STFC ERF fellowship (grant ST/N00485X/1) and HPC from the DiRAC facility (ST/P002293/1). 
M-MML was partly funded by NSF grant AST18-15461.  
R.T. was supported by DFG SFB 881 ``The Milky Way System'' and the Excellence Cluster STRUCTURES (EXC 2181-390900948), as well as ERC Synergy Grant ECOGAL (project ID 855130). 
The Cosmic Dawn Center of Excellence is funded by the Danish National Research Foundation under grant No. 140.

%

\vspace{5mm}


\software{\python version 3.7 (available at \url{http://www.python.org}),
  Astropy \citep{astropy:2013,astropy:2018},
  \arepo \citep{Springel2010},
  \cloudy version 17.2 \citep{Ferland17a}, 
  \skirt version 9.0 \citep{Camps2020},
  \swiftsimio \citep[\url{https://github.com/SWIFTSIM/swiftsimio}; ][]{Borrow2020})
}




\clearpage
\bibliography{bibs}

\clearpage
\appendix

\section{Properties of \simba-100 model galaxy sample}\label{app:100}

Table\,\ref{tab:sample1} lists the physical properties of the \simba-100 galaxy sample while Table\,\ref{tab:sample2} lists their line luminosities, including a few (\Loib, \Lcia and \Lcib) not used in the validation of the method.

\begin{deluxetable}{llllll}[h!]
  \tablecaption{Physical Properties of the Sample of \nsam \simba-100 Galaxies Used in This Work.
    \label{tab:sample1}}
\tablehead{\colhead{Galaxy}&\colhead{M$_\star$}&\colhead{M$_{\mathrm gas}$} &\colhead{SFR}& \colhead{\Zsfr}&\colhead{\Zmw} \cr \colhead{Index}&\colhead{($10^{9}$\,\Msun)}&\colhead{($10^{9}$\,\Msun)}&\colhead{(\uSFR)}&\colhead{(\Zsun)}&\colhead{(\Zsun)}}
\centering
\tablecolumns{6}
\startdata
0 & 0.4148 & 6.6530 & 0.35 & 0.16 & 0.02 \\ 
1 & 0.4594 & 7.6858 & 0.37 & 0.10 & 0.03 \\ 
2 & 0.4923 & 6.1911 & 0.18 & 0.12 & 0.05 \\ 
3 & 0.5001 & 9.5347 & 0.18 & 0.13 & 0.02 \\ 
4 & 0.5318 & 8.6812 & 0.38 & 0.16 & 0.04 \\ 
5 & 0.5342 & 6.3430 & 0.35 & 0.14 & 0.05 \\ 
6 & 0.5406 & 10.0238 & 0.54 & 0.15 & 0.04 \\ 
7 & 0.5561 & 7.0146 & 0.17 & 0.15 & 0.03 \\ 
8 & 0.5646 & 6.3235 & 0.52 & 0.11 & 0.02 \\ 
9 & 0.5948 & 5.8621 & 0.69 & 0.15 & 0.06
\enddata
\tablecomments{\Zmw stands for mass-weighted gas-phase metallicity. This table is available online in its entirety in machine-readable format.}
\end{deluxetable}

\begin{rotatetable*}
\begin{deluxetable*}{lccccccccccc}
\tablecaption{Line Luminosities for the Sample of \nsam \simba-100 Galaxies Used in This Work. \label{tab:sample2}}
\tablehead{\colhead{Galaxy}&\colhead{\Lcii}&\colhead{\Lcia}&\colhead{\Lcib}&\colhead{\LniiA}&\colhead{\LniiB}&\colhead{\Loia}  &\colhead{\Loib}&\colhead{\Loiii}&\colhead{\LcoA}&\colhead{\LcoB}&\colhead{\LcoC} \cr \colhead{index}&\colhead{(\Lsun)}&\colhead{(\Lsun)}&\colhead{(\Lsun)}&\colhead{(\Lsun)}&\colhead{(\Lsun)}&\colhead{(\Lsun)}  &\colhead{(\Lsun)}&\colhead{(\Lsun)}&\colhead{(\Lsun)}&\colhead{(\Lsun)}&\colhead{(\Lsun)}}
\tablecolumns{12}
\startdata
0 & 1.32e+07 & 4.38e+04 & 1.45e+05 & 2.73e+06 & 5.25e+06 & 3.39e+05 & 8.33e+06 & 4.10e+06 & 2.64e+02 & 5.82e+03 & 2.81e+04 \\ 
1 & 4.53e+07 & 2.33e+04 & 1.30e+05 & 1.38e+07 & 3.28e+07 & 1.35e+06 & 4.54e+07 & 3.65e+07 & 1.12e+02 & 3.15e+03 & 1.98e+04 \\ 
2 & 2.13e+07 & 5.74e+04 & 1.70e+05 & 4.22e+06 & 8.29e+06 & 1.16e+06 & 3.51e+07 & 5.67e+06 & 8.64e+02 & 2.20e+04 & 1.21e+05 \\ 
3 & 2.16e+07 & 5.54e+04 & 2.22e+05 & 5.43e+06 & 9.33e+06 & 5.15e+05 & 1.20e+07 & 2.95e+06 & 4.58e+02 & 1.13e+04 & 6.14e+04 \\ 
4 & 6.18e+07 & 3.01e+04 & 1.61e+05 & 1.83e+07 & 4.49e+07 & 2.54e+06 & 9.40e+07 & 1.16e+08 & 1.58e+02 & 4.47e+03 & 2.84e+04 \\ 
5 & 6.26e+07 & 3.20e+04 & 1.70e+05 & 1.85e+07 & 4.28e+07 & 1.96e+06 & 5.99e+07 & 4.45e+07 & 2.14e+02 & 6.08e+03 & 3.87e+04 \\ 
6 & 6.42e+07 & 5.20e+04 & 2.73e+05 & 1.93e+07 & 4.17e+07 & 1.79e+06 & 4.98e+07 & 3.76e+07 & 3.74e+02 & 9.91e+03 & 5.76e+04 \\ 
7 & 8.17e+06 & 3.89e+04 & 1.02e+05 & 1.38e+06 & 2.19e+06 & 2.94e+05 & 8.91e+06 & 8.28e+05 & 4.73e+02 & 1.08e+04 & 5.43e+04 \\ 
8 & 2.28e+07 & 3.07e+04 & 1.28e+05 & 4.43e+06 & 8.29e+06 & 7.76e+05 & 2.17e+07 & 3.89e+06 & 1.73e+02 & 4.68e+03 & 2.81e+04 \\ 
9 & 6.87e+07 & 3.20e+04 & 1.71e+05 & 1.82e+07 & 4.33e+07 & 3.87e+06 & 1.41e+08 & 1.10e+08 & 2.60e+02 & 7.38e+03 & 4.70e+04
\enddata
\tablecomments{This table is available online in its entirety in machine-readable format.}
\end{deluxetable*}
\end{rotatetable*}

\clearpage
\onecolumngrid

\section{Convergence tests with \skirt}\label{app:skirt}
\begin{figure}[htbp]
\centering
\includegraphics[width=0.45\textwidth]{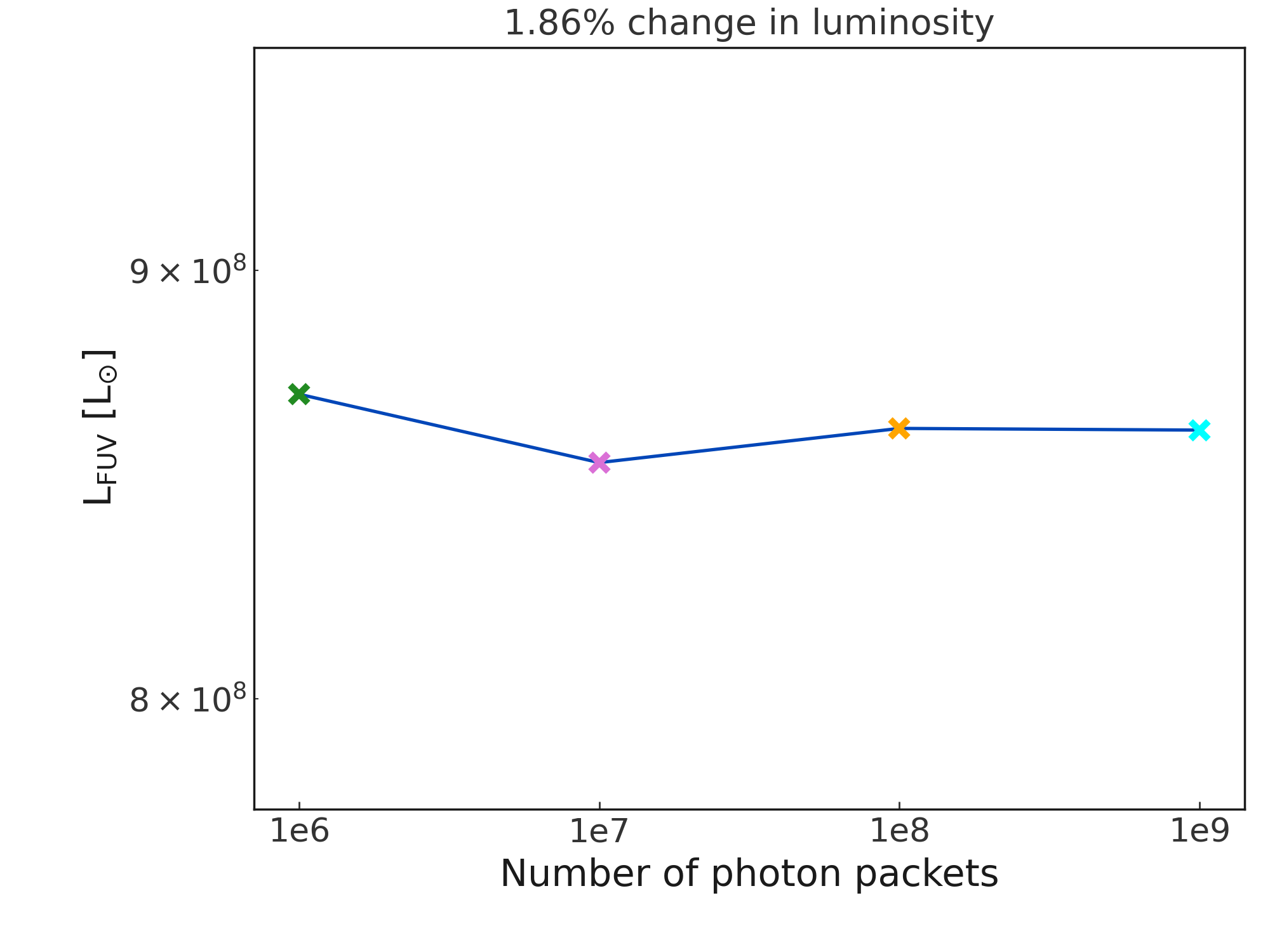}
\includegraphics[width=0.45\textwidth]{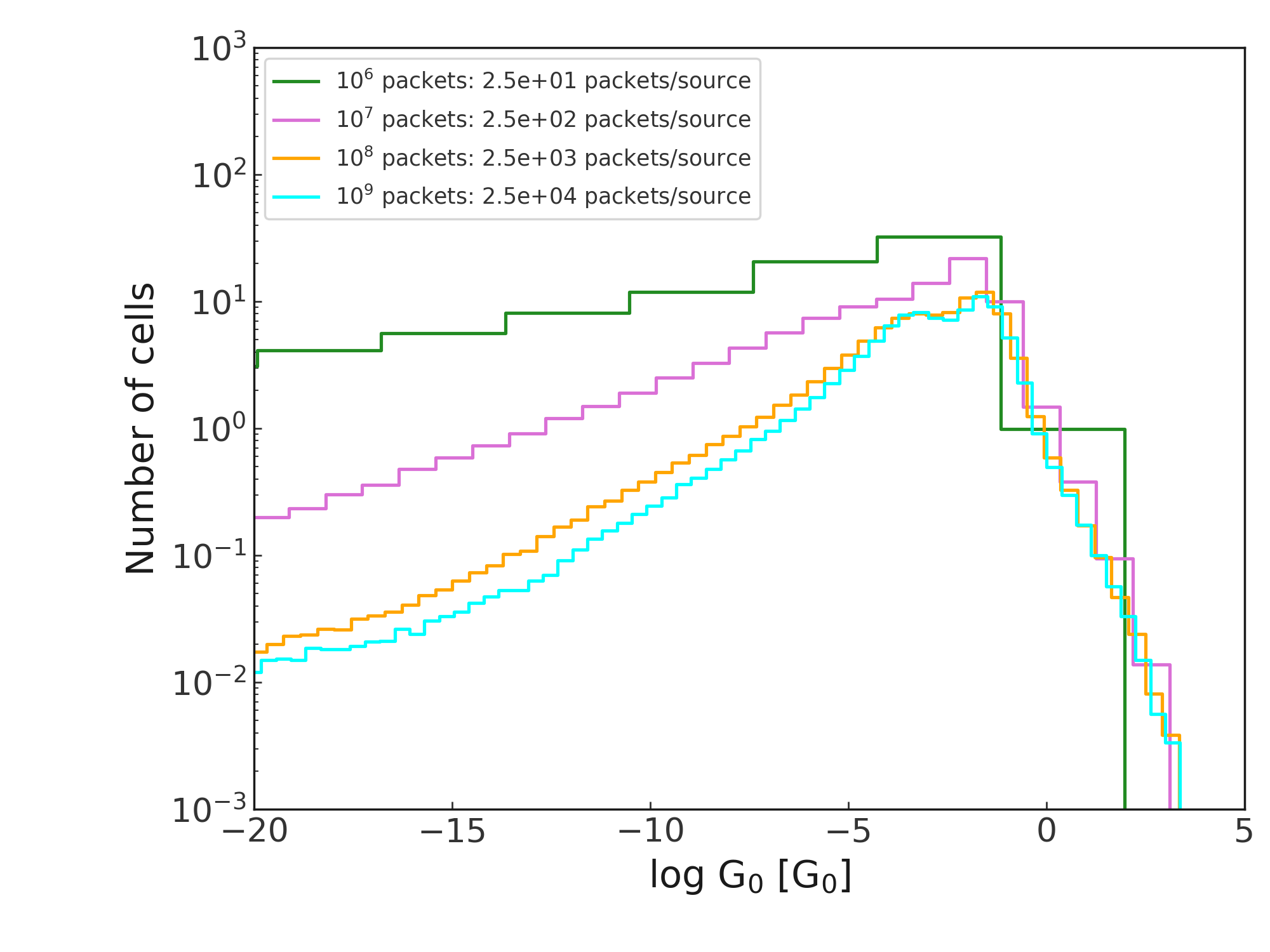}
\caption{Convergence tests of the \skirt FUV output for different total numbers of photon packets used in the calculation. \label{fig:app-skirt}}
\end{figure}

\skirt uses a fixed set of photon packets when iterating for a solution to the radiative transfer problem. 
We tested \skirt with different photon packet numbers for the galaxy with the largest stellar mass in \simba-100. corresponding to 39,733 stellar particles. 
The resulting total FUV luminosity and FUV flux distribution can be seen in Figure\,\ref{fig:app-skirt} for photon packet sizes of $10^6$, $10^7$, $10^8$ and $10^9$, corresponding $2.5\times10^1$, $2.5\times10^2$, $2.5\times10^3$ and $2.5\times10^4$ to photon packets per source. 
There is only a negligible change in FUV luminosity of $1.86\%$ compared to using the lowest number of photon packets, but when looking at the flux distribution it becomes clear that at least $10^8$ packets are necessary for a stable result, whereas increasing the number to $10^9$ has little effect. We therefore settled on $10^8$ photon packets.

\clearpage

\section{Distribution of cell physical parameters}\label{app:cell}

\begin{figure}[htbp]
\centering
\includegraphics[width=0.6\columnwidth]{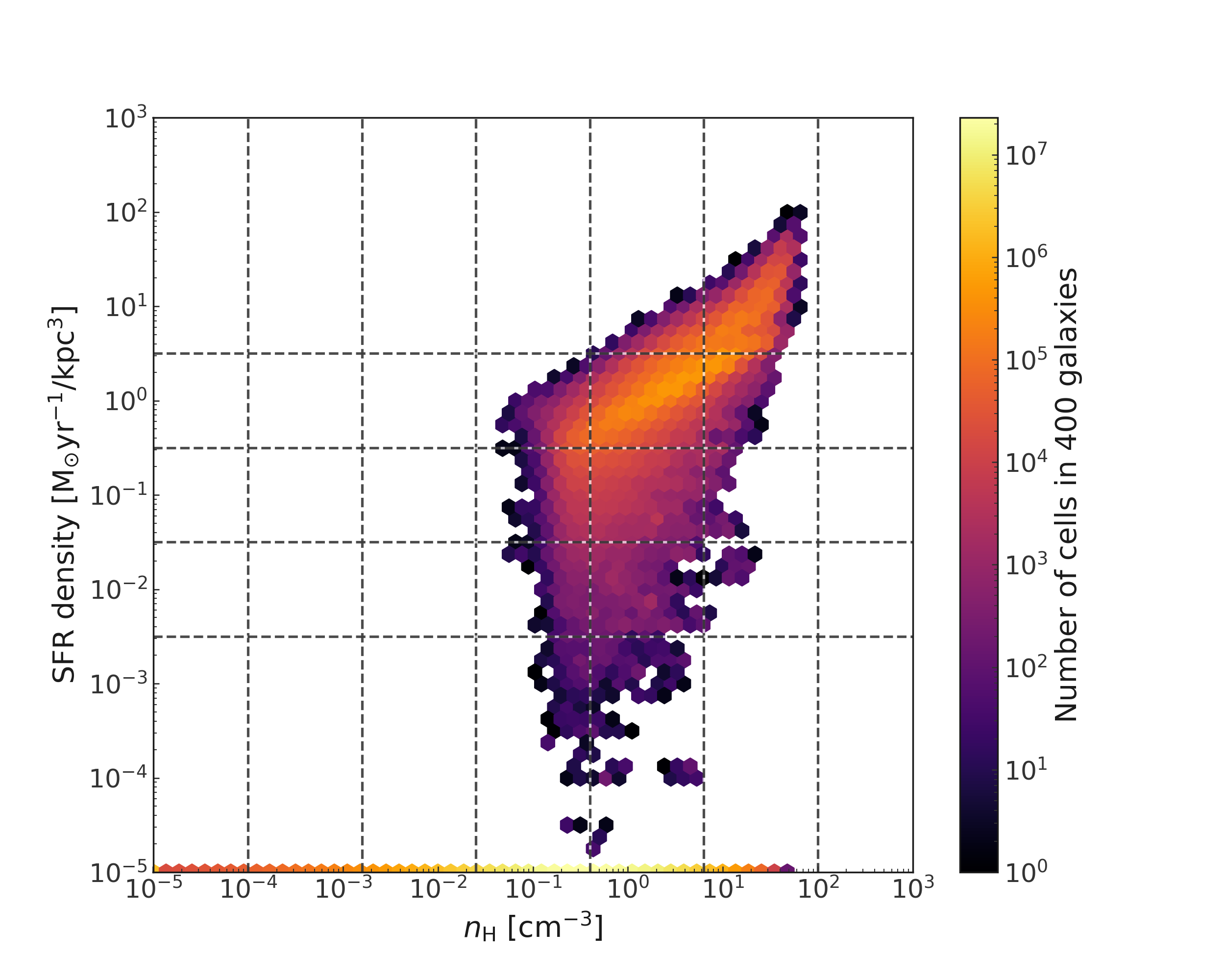}
\caption{Contour plot showing the cell distribution in density and SFR volume density for all \nsam sample galaxies. The horizontal dashed lines indicate the center of the \nSFR bins shown in Figure\,\ref{fig:AREPO_PDFs}, while the vertical dashed lines correspond to the densities used to sample the \simba gas densities in Section\,\ref{sec:step4}. Cells with \nSFR$=0$\,\unSFR are set to \nSFR$=10^{-5}$\,\unSFR in order to show their density distribution in the plot. \label{fig:cell}}
\end{figure}

Figure\,\ref{fig:cell} shows the volume-averaged gas and SFR densities for all cells in all galaxies of our sample. The grid point values in \nvw and \nSFR used to sample the \cloudy grid are also shown in Figure\,\ref{fig:cell}. The cells in the \simba galaxies have a larger spread in \nSFR and go to lower \nvw than the chosen grid points, reflecting a larger parameter space than what is found in \arepo-M51. However, we do not expect this to be a problem since the effect of \nSFR on the density PDF is less than that of \nvw as seen in Figure\,\ref{fig:AREPO_PDFs} and the regions of low \nvw are treated separately.

\clearpage

\section{Comparison of similar galaxies in \simba-100 and \simba-25}\label{app:25-100}

\begin{figure}[htbp]
\centering
\includegraphics[trim=80 0 80 0, clip, width=1.\textwidth]{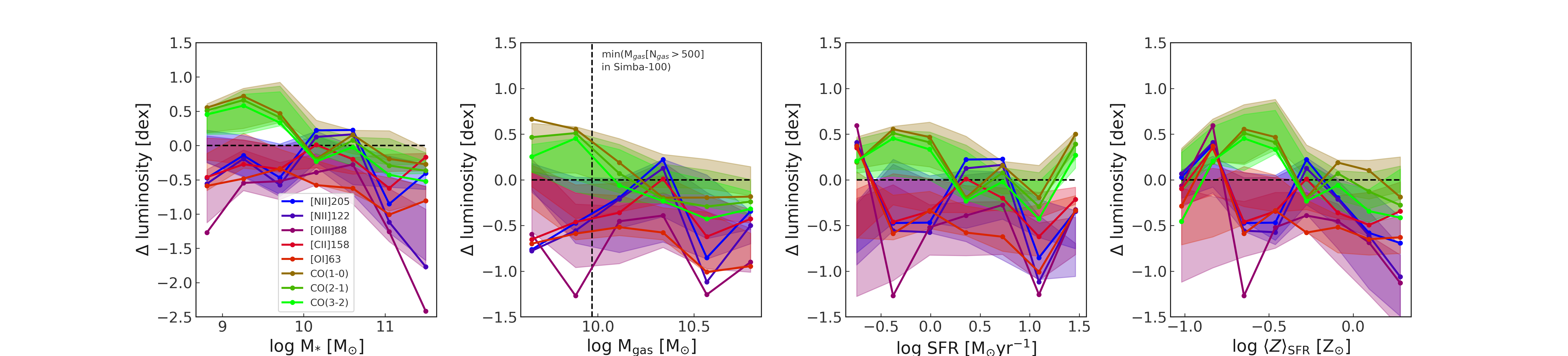}
\caption{Deviation in line luminosity between galaxies of similar \Mstar, \Mgas, SFR and \Zsfr in \simba-100 vs \simba-25 as a function of those global properties. Only the eight emission lines considered in this paper are considered, shown here with different colors. Shaded regions show the range in luminosities from the 25th to 75th percentile for all galaxy pairs in that bin.  \label{fig:25-100}}
\end{figure}

In order to make a fair comparison of the line emission calculated by \sigame for the different \simba volumes used here, we have selected a handful of galaxies in both \simba-100 and \simba-25 that have similar global properties in terms of \Mstar, \Mgas, SFR, and \Zsfr. 
We identify ``galaxy pairs'' by searching for the smallest distance in the 4D parameter space spanned by the \Mstar, \Mgas, SFR, and \Zsfr values, all in log units and normalized to lie in the range from 0 to 1. 
The result can be seen in Figure\,\ref{fig:25-100} in which the luminosity in \simba-100 is subtracted from the luminosity in \simba-25 for each galaxy pair to give a deviation in luminosity, as function of \Mstar, \Mgas, SFR and \Zsfr. 
The colored lines show the closest pair for the parameter of that panel while the shaded regions show the range in luminosities from the 25th to 75th percentile for all galaxy pairs in that bin. 
A dashed horizontal line shows a deviation of 0, signalling that the two galaxies have equal line luminosity. 
The luminosities of \niiA, \niiB, \oiii, \cii and \oi tend to be overestimated by up to $\sim0.5$--$1.5$\,dex in \simba-25 compared to \simba-100, in particular at higher values of SFR and \Zsfr. 
We attribute the larger deviations at low values of \Mstar and \Mgas to the lower mass resolution in \simba-100 compared to \simba-25. 
To illustrate the lack in resolution, a vertical dashed line indicates the minimum gas mass in our \simba-100 sample with at least 500 gas particles, which is $5.6\times10^9$\,\Msun. The mass resolution of \simba-25 is a factor of 8 higher, with the initial gas particle mass in \simba-100 being $1.82\times10^7$ \Msun, compared to $2.28\times10^6$ \Msun in \simba-25. Our \simba-25 selection contains galaxies with gas masses down to $7.3\times10^8$\,\Msun, with the constraint that they contain at least 500 gas particles.

\clearpage

\section{Impact of Different Density PDFs on Line Luminosity distribution}\label{app:line_nH}
\begin{figure}[htbp]
\centering
\includegraphics[width=0.4\textwidth]{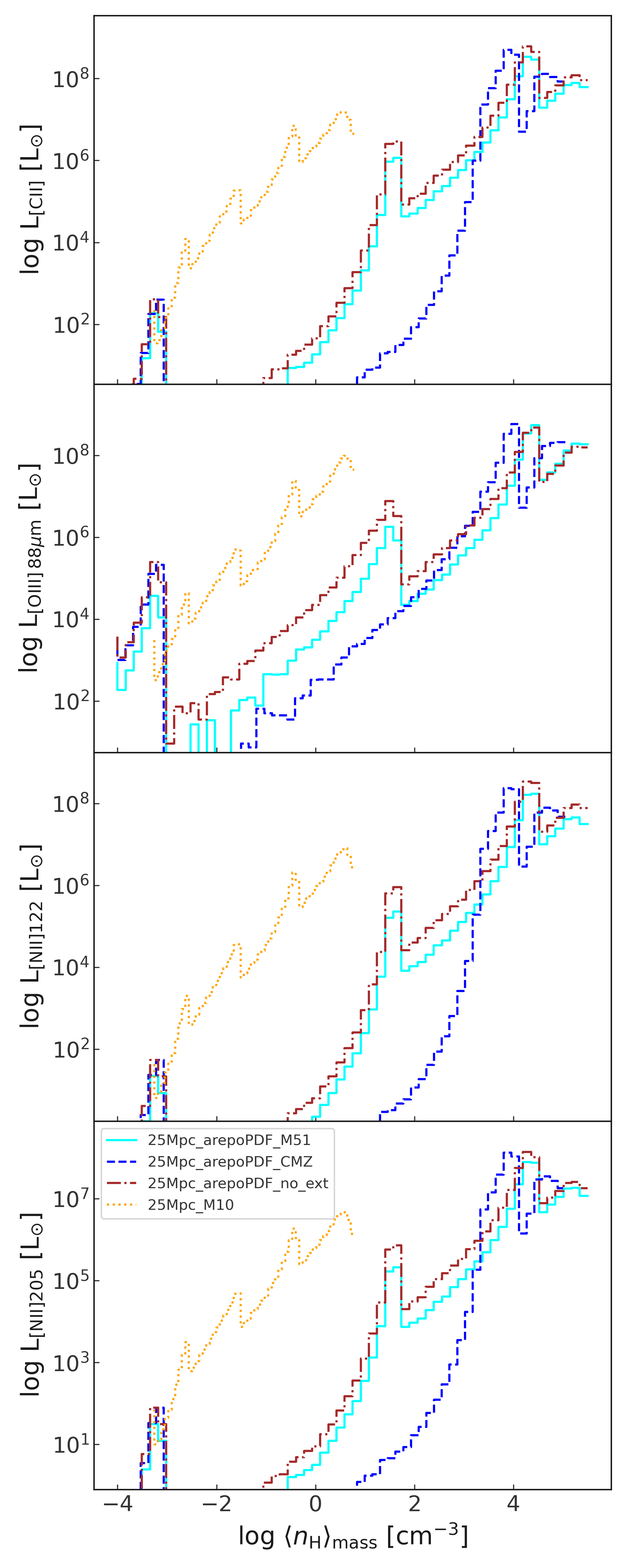}
\includegraphics[width=0.4\textwidth]{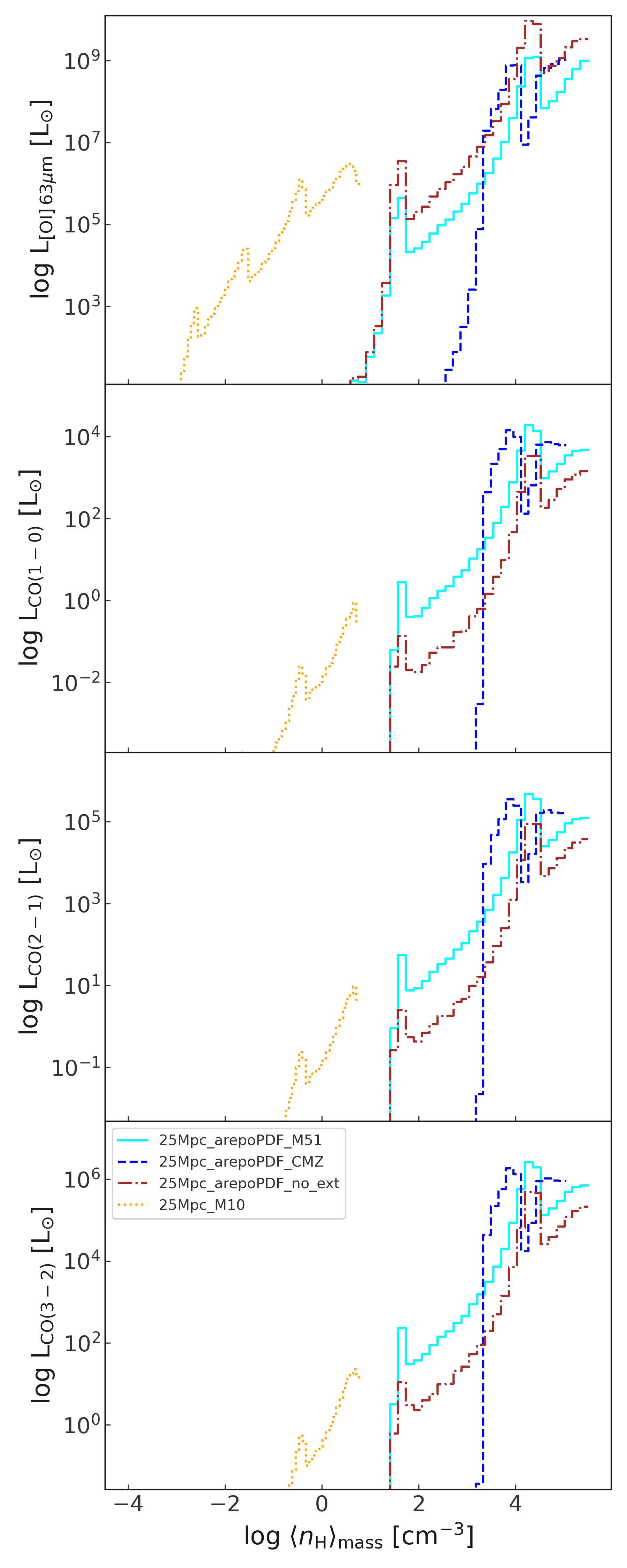}
\caption{Line luminosity per cell integrated over different hydrogen density bins in the \simba-25 galaxy shown in the top panels of Figs.\,\ref{fig:maps} and \ref{fig:mom0}. Different line styles refer to different methods for deriving the sub-grid density PDF: with the default \arepo-M51 simulation (cyan, solid), with the \arepo-CMZ simulation (blue, dashed), with the \arepo-M51 simulation and no attenuation at high densities (brown, dotted--dashed), and with a fixed lognormal shape corresponding to a Mach number of 10 and $b=1/3$, see eq.\,\ref{eq:Mach} (orange, dotted).   \label{fig:line_nH}}
\end{figure}

In Fig.\,\ref{fig:line_nH} we explore the origin of line emission with respect to hydrogen density for the different sub-grid density PDF prescriptions adopted in this paper for each of the eight emission lines investigated here. As shown in Fig.\,\ref{fig:ex_PDFs} the resulting sub-grid densities are generally no larger than around $10$\,\cc when restricting the density PDF shape to that of a lognormal of Mach number 10 and $b=1/3$ centered at the gas density from the simulation for each cell. In Fig.\,\ref{fig:line_nH} this means all line luminosity is restricted to come from densities below around $10$\,\cc, as the orange dotted lines show. In Section\,\ref{sec:CMZ} we describe the \arepo-CMZ run, which was used as an alternative simulation to generate tables of density PDFs. In particular, the PDF tables generated with \arepo-CMZ yield more mass at higher densities in the regime of low average cell density ($\log\nH < 0$). This can also be seen in Fig.\,\ref{fig:line_nH}, where the line luminosity distributions with \arepo-CMZ (blue, dashed) are skewed toward higher densities than the default result using \arepo-M51 (cyan, solid). Due to the differences between the M51 case and the comparatively extreme CMZ environment and hence between the two simulations as described in Section\,\ref{sec:CMZ}, these significant changes in line luminosity distribution are indeed expected. Finally, the run using the \arepo-M51 simulation but no attenuation at high densities (brown, dotted--dashed), as otherwise done following the description in Section\,\ref{sec:step4}, returns the same density distribution as the default run, but the missing extinction results in overall lower or higher luminosities, depending on the line considered. Specifically, for the higher ionization lines of \niiA, \niiB and \oiii the line luminosities are higher without the extinction in the FUV, while the CO line luminosities are lower, since the missing extinction means that less CO is able to form. 
A note of caution when interpreting Fig.\,\ref{fig:line_nH} is that the density on the x-axis is a mass-weighted density of all the one-zone \cloudy models used to calculate the line luminosity of that cell. It is not weighted by where the line emission is actually coming from (luminosity-weighted).

\end{document}